\documentclass[aps,prd,amssymb,eqsecnum,showpacs]{revtex4}
\usepackage{graphicx}
\usepackage{psfrag}
\usepackage[T1]{fontenc}          
      
\usepackage[latin1]{inputenc}       
\usepackage{latexsym}                 
\usepackage{moreverb}                                                    
\usepackage{textcomp}                
\usepackage{subfigure}
\usepackage{epsfig}                
\usepackage{color}
\usepackage{amsfonts}
\usepackage{amsmath}                 
\usepackage{amsthm}
\begin{document}
\title{Finite difference schemes for second order systems describing black
holes}

\author{Mohammad Motamed${}^{1,2}$, M. Babiuc${}^{2,3}$, B. Szil\'{a}gyi${}^{2}$,
	H-O. Kreiss${}^{1,2}$, 
	 and
	J. Winicour${}^{2,3}$}
\affiliation{
${}^{1}$NADA, Royal Institute of Technology, 10044 Stockholm, Sweden\\
${}^{2}$Albert Einstein Institute, Max Planck Gesellschaft,
Am M\"uhlenberg 1, D-14476 Golm, Germany\\
${}^{3}$Department of Physics and Astronomy,
University of Pittsburgh, Pittsburgh, Pennsylvania 15260
}

\date {Mar. 19}

\begin{abstract}

In the harmonic description of general relativity, the principle part of
Einstein's equations reduces to 10 curved space wave equations for the
components of the space-time metric. We present theorems regarding the
stability of several evolution-boundary algorithms for such equations when
treated in second order differential form. The theorems apply to a model
black hole space-time consisting of a  spacelike inner boundary
excising the  singularity, a timelike outer boundary and a horizon in between.
These algorithms are implemented as stable, convergent numerical codes and
their performance is compared in a 2-dimensional excision problem.

\end{abstract}


\maketitle
\section{Introduction}

A primary goal of numerical relativity is the computation of
gravitational radiation waveforms from binary black holes. Radiation
produced in the inspiral and merger of binary black holes is expected
to provide a strong signal for gravitational wave observatories.
However, the simulation of black holes has proved to be a difficult
computational problem. The importance of this challenging problem has
recently spurred a fertile interaction between numerical relativity
and computational mathematics. The classic computational treatment of
hyperbolic systems has been directed at fluid dynamics and has been
based upon first differential order systems. Certain formulations of
Einstein's equations take a more natural second order form, notably
the harmonic formulation~\cite{dedonder,Fock} for which well-posedness
of the Cauchy problem was first established~\cite{Choquet}. Here we
present theorems regarding the stability of several evolution-boundary
algorithms for such second order systems which have direct application
to the black hole problem.

Harmonic coordinates $x^\alpha=(t,x^i)=(t,x,y,z)$ have only recently
been used in designing numerical
codes~\cite{garf,szilschbc,harl,bab,pret,pret2,babev,linsch,harm}.
They satisfy the curved space wave equation
\begin{equation}
      \Box_g x^\mu :=  \frac{1}{\sqrt{-g}}\partial_\alpha
            (\sqrt{-g}g^{\alpha\beta}\partial_\beta \, x^\mu ) =0.
\end{equation}
In harmonic coordinates, Einstein's equations reduce to 10 quasilinear
wave equations for the components of the metric,
\begin{equation}
      \Box_g g^{\mu\nu} = S^{\mu\nu} ,
             \label{eq:nlg}
\end{equation}
where $S^{\mu\nu}$ are nonlinear terms which do not enter the
principle part. Thus the scalar wave equation
\begin{equation}
      g^{\alpha\beta} \partial_\alpha \partial_\beta  u  =0,
      \label{eq:3wave}
\end{equation}
which has the same principle part, provides a fundamental testing
ground for designing algorithms to treat the nonlinear gravitational
problem (\ref{eq:nlg}). In a previous study~\cite{excis}, we used this
scalar equation to develop evolution and boundary algorithms for a
model one dimensional black hole excision problem. Here we extend
these results to two dimensions. While the extension to 2D involves
substantial new features, the generalization from 2D to 3D is quite
straightforward. Thus our results are immediately applicable to
algorithms for the harmonic gravitational equations (\ref{eq:nlg}), as
well as their generalization to include harmonic gauge forcing
terms~\cite{Friedrich} and other related generalizations such as the
$Z4$ formulation~\cite{z4}. 

We treat (\ref{eq:3wave}) in the second order differential form, which
has advantages for both computational efficiency and accuracy over
first order formulations~\cite{krsecor,kreissort}. Although the system
can be reduced to first order symmetric hyperbolic form~\cite{fisher},
this has the disadvantage of introducing auxiliary variables with
their associated constraints and boundary conditions. The second order
form is also best suited to the analogous wave equations governing
elasticity and acoustics. Elasticity theory is governed by a coupled
system of wave equations which for simple cases is similar to
(\ref{eq:3wave}), in which the spatial components $g^{ij}$ are
determined by the elastic moduli. In fact, some of the techniques
utilized here have been developed in a recent computational study of
the wave equations governing an elastic body~\cite{krneum}. The new
ingredient introduced in the wave equation (\ref{eq:3wave}) arises from the
non-vanishing mixed space-time derivatives arising from the components
$g^{it}$. Such terms do not ordinarily appear in the wave equations
governing elasticity theory because they are treated in the rest frame
of the body but they would necessarily arise in treating acoustic
waves propagating in a medium with nonuniform macroscopic motion. In
general relativity, these mixed space-time components of the metric
correspond to a non-vanishing ``shift'', which is an essential feature
of the black hole problem. In the standard $3+1$ description of
space-time~\cite{adm}, the Cauchy hypersurfaces $t=constant$ are
required to be spacelike so that they have a length element with
Euclidean signature
\begin{equation}
  d\ell^2= h_{ij}dx^i dx^j.
\end{equation}
The inverse spatial metric $h^{ij}$, satisfying $h^{ij}h_{jk}
=\delta^i_k$, is related to the spatial components of the 4-metric
determining the wave operator by
\begin{equation}
   h^{ij}=g^{ij}-\frac{1}{g^{tt}}\beta^i \beta^j ,
\end{equation}
where $\beta^i$ are the components of the shift. Here, the spacelike
character of the Cauchy hypersurfaces requires that $g^{tt}<0$.

The wave equation with shift has not received a great deal of
attention outside of recent work in general
relativity~\cite{alcsch,harl,calab,excis,calabgund,babev}. Even before
the computational problem is attempted, new mathematical features
introduced by the shift must be incorporated into the formulation of a
well-posed initial-boundary value problem. The operator
$h^{ij}\partial_i\partial_j$ is by construction an elliptic operator
defined by the spatial metric of the Cauchy hypersurfaces. However,
the operator $g^{ij}\partial_i\partial_j$ is elliptic only when the
shift is sufficiently small. The elliptic case arises when the
operator $\partial_t$ is timelike, i.e. when the evolution proceeds in
a timelike (subluminal) spacetime direction. 

Without loss of generality, we set $g^{tt}=-1$ and write the 2D
version of (\ref{eq:3wave}) as
\begin{equation}
  \bigg ( \partial_t^2 - 2(\beta^x \partial_x +\beta^y \partial_y)\partial_t 
      - (a_1-\beta^x\beta^x)\partial_x^2 -(c_1-\beta^y\beta^y) \partial_y^2
       -2(b_1-\beta^x \beta^y)\partial_x \partial_y \bigg )u=0,
      \label{eq:2wave}
\end{equation}
where $h^{xx}=a_1$, $h^{xy}=b_1$ and $h^{yy}=c_1$ .
The Euclidean property of $h^{ij}$ requires 
\begin{equation}
       a_1>0 \, , \quad c_1>0 \, , \quad a_1 c_1-b_1^2 >0.
\end{equation}
The components of $g^{ij}$ are $g^{xx}=a=a_1-\beta^x\beta^x$,
$g^{yy}=c=c_1-\beta^y\beta^y$ and $g^{xy}=b=b_1-\beta^x\beta^y$.  In the
subluminal case when $g^{ij}\partial_i \partial_j$ is an elliptic
operator, the simplest second order accurate difference approximation
to (\ref{eq:2wave}) is
\begin{equation}
    W:=\bigg (\partial_t^2 -2(\beta^x D_{0x} +\beta^yD_{0y} )\partial_t 
         -aD_{+x} D_{-x}- c D_{+y} D_{-y} -2 b D_{0x} D_{0y}
	 \bigg) u   =0.
      \label{eq:subwave}
\end{equation}
(Here $D_{0i}$, $D_{+i}$ and $D_{-i}$ are, respectively, the centered,
forward and backward difference operators in the $x^i$-direction
defined in Sec.~\ref{sec:algorithms}). This leads to stable
evolution-boundary algorithms for Dirichlet, Neumann, Sommerfeld or
other dissipative boundary conditions. Stability was established for
the 1D case using a semi-discrete energy norm in~\cite{excis}, and
this was generalized using the discrete energy method to the full 3D
case in~\cite{babev,harm}.

The $W$-algorithm (\ref{eq:subwave}) is unstable when the shift is 
sufficiently large so that $g^{ii} \le 0$ (for any diagonal component).
This occurs in one of the strategies for avoiding problems with the
singularity which ultimately forms inside a black hole. In this
strategy, the singularity is ``excised'' by surrounding it with a
spacelike inner boundary. The evolution direction which is adopted to
this spacelike boundary is superluminal, so that $g^{ij}$ no longer has
Euclidean signature. In Sec.~\ref{sec:algorithms}, we establish the
stability of several different second order, evolution-boundary
algorithms for this superluminal case.  For the Cauchy problem, we
establish stability for a general system of wave equations in $s$
spatial dimensions so that the results may be immediately applied to
other second order systems such as elasticity theory or acoustics. For
the boundary, we specialize our treatment to scalar equations in $2$D in
order to simplify the notation, but the extension to general systems in
$s$D is straightforward. In particular, our results apply directly to
the 3D harmonic evolution of black holes.

The analysis of the initial-boundary problem for (\ref{eq:2wave})
in Sec.~\ref{sec:algorithms} makes evident that the above geometric
properties of the wave equation have a mathematical analogue which
results independently from a consideration of the well-posedness of
the problem. The geometrical and analytical approaches are
complementary and provide a good meeting ground for the ideas of
numerical relativity and computational mathematics. While the main
concern of numerical relativity is the black hole problem, the
stability theorems for the finite difference algorithms developed
for the model problems considered here provide a firm basis for
attacking this problem with the harmonic Einstein system
(\ref{eq:nlg}). 

In Sec.~\ref{sec:periodic}, we compare the performance of the algorithms
for the superluminal case in a problem without boundaries. In
Sec.~\ref{sec:excision}, we simulate a simple 2D model of the excision
problem in which the inner boundary ${\cal S}$ is spacelike and the
outer boundary ${\cal T}$ is timelike. Between the boundaries the
operator $g^{ij}\partial_i\partial_j$ goes from non-elliptic to elliptic
along a curve ${\cal H}$ where $\det(g^{ij})=0$. The metric is chosen so
that no characteristics can leave the inner region between ${\cal S}$
and ${\cal H}$, so that ${\cal H}$ mimics the role of a horizon. The
global simulation of (\ref{eq:2wave}) in the region bounded by ${\cal
T}$ and ${\cal S}$ is achieved with a blended evolution algorithm. A
stable superluminal algorithm is used in an inner region between ${\cal
S}$ and ${\cal H}$. In the exterior region, this superluminal algorithm
is blended to the $W$-algorithm (\ref{eq:subwave}), so that the
$W$-algorithm is used to treat the outer boundary ${\cal T}$. This model
excision problem involves many of the mathematical difficulties in the
full gravitational case. We begin in Sec.~\ref{sec:subtle} with some
simple examples which illustrate the problem, its potential pitfalls 
and how to avoid them. 

\section{Some subtleties associated with the wave equation with shift}
\label{sec:subtle}

In an inertial coordinate system $\hat x^\alpha=(\hat t,\hat x^i)$ (in
units where the velocity of light $c=1$), the wave equation which
governs special relativistic physics,
\begin{equation}
   \bigg (\partial_{\hat t}^2 -\delta^{ij} 
      \partial_{\hat i} \partial_{\hat j}\bigg )u =0,
    \label{eq:wnoshift}
\end{equation}
does not contain a shift. The invariance of the velocity of light
results from the property that this wave equation retains the same
form (\ref{eq:wnoshift}) under a Lorentz transformation,
\begin{eqnarray}
  t' &=& \frac{1}{\sqrt{1-\beta^2}}
      ( \hat t-\delta_{ij}\beta^i  \hat x^j) \nonumber \\
 x'^i &=& \frac{1}{\sqrt{1-\beta^2}}
      ( \hat x^i - \beta^i  \hat t) \nonumber \\
      \beta^2 &=&\delta_{kl}\beta^k\beta^l ,
\end{eqnarray}
to another inertial coordinate system with relative motion. In this
way special relativity resolves the dilemma with experiment that under
a Galilean transformation,
\begin{eqnarray}
  t &=&\hat t \nonumber \\
  x^i &=& ( \hat x^i - \beta^i  \hat t) ,
  \label{eq:gal}
\end{eqnarray}
(\ref{eq:wnoshift}) gives rise to the shifted wave equation
\begin{equation}
    \bigg (\partial_{t}^2 - 2\beta^i\partial_i
    -(\delta^{ij}-\beta^i \beta^j)\partial_i \partial_j
       \bigg )u =0
    \label{eq:wshift}
\end{equation}
whose solutions propagate with coordinate speeds in the range $|1 \pm
\beta|$ (where $\beta^2=\delta_{ij}\beta^i \beta^j$). This raises the
question: why does the wave equation with shift arise in general
relativity?

In fact, although there are no preferred inertial coordinates in
general relativity, in any sufficiently small spacetime region it is
always possible to introduce Gaussian coordinates in which the wave
equation (\ref{eq:3wave}) reduces to the shift-free form 
\begin{equation}
   ( \partial_t^2 - h^{ij} \partial_i \partial_j ) u  =0.
\end{equation}
The problem here is that in Gaussian coordinates the worldlines
$x^i=constant$ are geodesics, i.e. the worldlines of freely falling
observers, which can be focused by the attractive nature of gravity to
produce coordinate singularities. This can occur on a short time scale
in a strong gravitational field.

Another reason for introducing a shift is the simplicity of harmonic
coordinates in reducing Einstein's equations into the hyperbolic form
(\ref{eq:nlg}). Since the shift components $g^{it}$ satisfy a coupled
system of nonlinear wave equations, even if they were initialized with
vanishing Cauchy data they would in general evolve to be non-zero.
This cannot be avoided by introducing a harmonic gauge forcing term,
of the form $\Box x^\alpha =F^\alpha$, without choosing the forcing
term $F^\alpha$ to depend upon the derivatives of the metric
$\partial_\mu g^{\alpha\beta}$. This in turn jeopardizes the
hyperbolic form of the reduced Einstein equations and the
well-posedness of the Cauchy problem~\cite{Friedrich}.

Another reason for introducing a shift arises in the simulation of
black holes. Once a black hole of mass $M$ has formed there is at most
a proper time of order $M$ (in gravitational units) until a physical
singularity is encountered. On the other hand, a simulation which
provides gravitational waveforms of physical interest typically
requires an evolution for a proper time of more than $100M$ in the
exterior region. One strategy for accomplishing this is to excise the
singularity by surrounding it with a spacelike inner boundary for the
simulation domain, i.e. an inner boundary which moves at superluminal
speed. If the evolution tracks the inner boundary then a superluminal
shift must be used.

This can be illustrated by a spherically symmetric Schwarzschild black
hole for which the wave equation (\ref{eq:3wave}) becomes
\begin{equation}
   \bigg ( (1+\frac{2M}{r}) \partial_t^2 -\frac{4M}{r} \partial_t \partial_r
           -(1-\frac{2M}{r}) \partial_r^2 
     -\frac{1}{r^2}( \partial_\theta^2+ \frac{1}{\sin^2 \theta}\partial_\phi^2)
   \bigg) u  =0,
   \label{eq:schwave}
\end{equation}
in ingoing Eddington-Finkelstein coordinates. Here the evolution takes
place on the spacelike Cauchy hypersurfaces $t=const$ which are
non-singular for $r>0$. The black hole is located at $r=2M$, which is
a characteristic hypersurface with the horizon property that no
characteristics leave the region $r\le 2M$. The singularity is excised
by evolving in a domain $R_1\le r \le R_2$, where $0<R_1<2M$ and
$R_2>> 2M$. The shift has the radial component 
\begin{equation}
  \beta^r =\frac{1}{1+\frac{r}{2M}}>0.
\end{equation} 
The change in sign of the coefficient of $\partial_r^2$ in
passing inside the horizon does not change the hyperbolicity of
the wave equation but it changes its mathematical properties.
Outside the horizon, the curves of constant $(r,\theta,\phi)$ 
are timelike, as well as the outer boundary $r=R_2$. In the
outer region $2M<r\le R_2$, the $W$-algorithm (\ref{eq:subwave})
provides a stable second order evolution-boundary algorithm for the wave
equation~\cite{excis,babev,harm}.

Inside the horizon, the $t$-direction, as well as the inner
boundary $r=R_1$ is spacelike, i.e. evolution on a grid with
constant $(r,\theta,\phi)$ proceeds outside the light cone. This
effects the mathematical properties of the wave equation. As a
result, in this domain, the $W$-algorithm is unstable. The alternative
algorithms presented in Sec.~\ref{sec:algorithms} are stable
inside the horizon. But the $W$-algorithm has better accuracy
than these algorithms in the exterior region~\cite{babev}. In
the simulation of the model excision problem in
Sec.~\ref{sec:excision}, a stable algorithm for the superluminal
regime is blended to the $W$-algorithm in the exterior.

The Schwarzschild horizon has the property that characteristics
can not exit from inside, but can enter from the outside. Near
the horizon, the radial part of Schwarzschild wave equation
(\ref{eq:schwave}) has the same qualitative features as the wave
equation 
\begin{equation}
   (\partial_t-\partial_x)(\partial_t+x\partial_x) u=0,
   \label{eq:xhor}
\end{equation} which has a horizon  $x=0$. One set of characteristics
of (\ref{eq:xhor}) cross the horizon at $x=0$ in the negative
$x$-direction. The other set of characteristics are tangent to the
horizon and {\em diverge} away on either side. An observer at $x>0$
cannot see beyond the horizon at $x=0.$   This is the situation which
we dealt with in a model 1D excision problem~\cite{excis} whose
treatment we generalize to 2D in Sec.~\ref{sec:excision}. However, it
should be emphasized that the related equation
\begin{equation}
   (\partial_t-\partial_x)(\partial_t-x\partial_x) u=0
   \label{eq:mxhor}
\end{equation}
has a different mathematical character. Although (\ref{eq:mxhor}) is
also hyperbolic and has a well-posed Cauchy problem, one set of
characteristics {\em converge} toward the horizon at $x=0$.  These
characteristics approach each other exponentially fast and, in
general, the gradients become exponentially large near $x=0$.  This
would lead to the focusing of a wave into formation of a shock.
Although we do not treat this case in this paper, it is important to
bear in mind that it would require different methods.

Boundaries introduce additional subtleties. First consider a
timelike boundary, similar to the outer boundary $r=R_2>2M$ for
the Schwarzschild wave equation (\ref{eq:schwave}). Since the
evolution is timelike in the neighborhood of the boundary, the
$W$-algorithm can be used. The stability of dissipative boundary
conditions for the $W$-algorithm was established for 1D
in~\cite{excis} and extended to 3D in~\cite{harm} by means of a
semi-discrete energy method. However, such an energy estimate does
not preclude exponential growth of a wave traveling between two
boundaries.  A simple example~\cite{bab} arises from the
repetitive blue shifting of a wave packet in special relativity
reflecting back and forth between two plane boundaries, whose
velocities $\pm v$ are controlled to be always toward the packet
during reflection. After many reflections the wave packet
shrinks in size and its energy grows by a factor $e^{4\alpha
T}$, where $T$ is measured in units of the crossing-time between
reflections and $v= \tanh \alpha$. Dissipation must be used to
control such growth of short wavelength error.

It is instructive to interpret the boundary conditions on a wave
in special relativity in the shifted coordinate system
(\ref{eq:gal}) where the boundary has fixed location but moves
relative to the  $t=const$ Cauchy hypersurfaces. In the 1D case,
this gives rise to the half-plane problem
\begin{equation}
    \bigg (\partial_{t}^2 - 2\beta\partial_x \partial_t
    -(1-\beta^2)\partial_x^2  \bigg )u =0,
    \label{eq:1wshift} 
\end{equation}
in the region $x\le 0$ (where we now write $\beta^x=\beta$). There
are two different frames in which the energy of the wave can be
considered - the rest frame of the boundary and the rest frame
intrinsic to the Cauchy hypersurfaces. In the rest frame of the
boundary, the energy is 
\begin{equation}
      E=\frac{1}{2}\int_{-\infty}^0 dx \bigg ( (\partial_t u)^2
      +(1-\beta^2)(\partial_x u)^2 \bigg )
\end{equation}
and satisfies
\begin{equation}
  \partial_t E= \partial_t u
      \bigg ((1-\beta^2)\partial_x+\beta \partial_t \bigg )u|_{x=0}.
      \label{eq:fluxcon}
\end{equation} In the case $\beta^2<1$, this energy provides a
norm  and the semi-discrete version of the flux-conservation law
( \ref{eq:fluxcon}) provides the basis for establishing stable
evolution-boundary algorithms for the $W$-algorithm
(\ref{eq:subwave}). Note the sign of $\beta$ is important here
in formulating a stable Neumann boundary
condition.
A homogeneous Neumann boundary
condition takes the dissipative form 
\begin{equation}
   \bigg ( (1-\beta^2)\partial_x+\beta \partial_t \bigg )u =0. 
\end{equation}
The familiar form $\partial_x u=0$ implies $\partial_t E \le 0$ and thus
guarantees a well-posed problem only when $\beta>0$, i.e. only when  the motion
of the boundary is outward relative to the Cauchy hypersurfaces.

The energy intrinsic to the Cauchy hypersurfaces,
\begin{equation}
      E_0=\frac{1}{2}\int_{-\infty}^0 dx \bigg (
         (\partial_t u-\beta \partial_x u)^2 
	 + (\partial_x u)^2 \bigg),
	 \label{eq:resten}
\end{equation}
provides a norm even in the superluminal case
when $\beta^2>1$. It satisfies
\begin{equation}
      \partial_t E_0= \bigg (
    \frac {\beta}{2} (\partial_t u -\beta \partial_x u)^2
    +\frac {\beta}{2} (\partial_x u)^2
    + (\partial_t u -\beta \partial_x u)\partial_x u \bigg ) |_{x=0}.
      \label{eq:rfluxcon}
\end{equation} 
Thus, in the absence of a boundary, (\ref{eq:rfluxcon}) would reduce to
$\partial_t E_0=0$ so that the Cauchy problem is well-posed for any
$\beta$. The energy analogous to $E_0$ is used in
Sec.~\ref{sec:algorithms} to establish well-posedness of the Cauchy
problem and the stability of superluminal algorithms in the general
multi-dimensional case.  

When $\beta<-1$, i.e. when the motion of the boundary is superluminal
and directed toward the Cauchy hypersurfaces, it is easy to verify that
(\ref{eq:rfluxcon}) implies $\partial_t E_0<0$ so that there is always a
loss of energy through the boundary. This is the case of a spacelike
boundary through which all the characteristics leave, i.e. a pure
``outflow'' boundary. Stable algorithms for such a boundary are also
given in Sec.~\ref{sec:algorithms} for the higher dimensional case. Note
that for $\beta>1$ the boundary is also spacelike but now 
(\ref{eq:rfluxcon}) implies $\partial_t E_0>0$. This is the pure
``inflow'' case, in which all the characteristics enter the boundary.
This should not be considered in the context of an initial-boundary
value problem, but as a pure Cauchy problem where the boundary
represents a non-smooth extension of the Cauchy hypersurface.

Further subtleties arise in treating co-orbiting, binary black holes. One
strategy for the binary problem is to use a rotating coordinate system which
co-orbits with the black holes. In the Schwarzschild case, the use of a
coordinate $\varphi=\phi-\omega t$ rotating with angular velocity $\omega$
transforms the wave equation (\ref{eq:schwave}) into 
\begin{equation}
   \bigg ( (1+\frac{2M}{r}) (\partial_t+\omega\partial_\varphi)^2 
       -\frac{4M}{r} \partial_t \partial_r -(1-\frac{2M}{r}) \partial_r^2 
   -\frac{1}{r^2}( \partial_\theta^2
   + \frac{1}{\sin^2 \theta}\partial_\varphi^2) \bigg) u  =0.
   \label{eq:coschwave}
\end{equation}
Now the $t$-direction becomes spacelike in the region 
\begin{equation}
      (1+\frac{2M}{r})r^2\omega^2 \sin^2 \theta >1,
\end{equation}
which intersects the outer boundary $r=R_2$ if $R_2$ is sufficiently large. In
that case, although the boundary remains timelike the evolution is
superluminal  so that the $W$-algorithm is no longer stable. A stable
algorithm for such a boundary problem has been established in the 1D
case~\cite{calabgund}. We will not consider the 2D version of this problem
here.

A common strategy for treating the binary black hole problem is to use a grid
based upon Cartesian coordinates. This poses a problem in dealing with inner
and outer boundaries with the spherical shapes natural to the problem. In other
second order wave problems, such curved boundaries have been successfully
treated by the embedded boundary method~\cite{krdir,krneum}. Another approach
being explored in general relativity is to use multi-block
grids~\cite{pfeif,schpfeif,sbp,mbs}. This is another problem which  we
defer to future work and do not consider here. 


\section{Algorithms for the 2D superluminal problem}
\label{sec:algorithms}

In this section, we study a class of second order hyperbolic systems
with shift which we will use in Sec.~\ref{sec:excision} to construct
stable algorithms for a model 2D black hole excision problem. The
excision problem is a strip problem with spacelike and timelike
boundaries and a horizon in between. In the region where the shift is
superluminal, the boundary is spacelike and where the shift is
subluminal, the boundary is timelike. We replace this problem by
Cauchy and half-space problems. The strip problem is well-posed if the
corresponding Cauchy and half-space problems are
well-posed~\cite{green-book}.

For the Cauchy problem, we consider general systems of equations in
$s$ space dimensions to demonstrate that the results have
applicability beyond numerical relativity. For the half-space
problems, we only consider scalar equations in 2D to simplify the
notation. The generalization from scalar equations in $2$D to systems
in $s$D is quite straightforward.

Here, we consider systems with constant coefficients. Systems with
variable coefficients can be reduced to systems with constant
coefficients by freezing the coefficients at all points. The problem
with variable coefficients is strongly well-posed if the Kreiss
condition holds uniformly for all problems with constant
coefficients~\cite{brown-book}.

In order to analyze and establish stable approximations we use the
method of lines and reduce the system of partial differential
equations to a system of ordinary differential equations in time on a
spatial grid. We then apply two standard techniques: the energy method
and mode analysis. The stability of the semi-discrete approximation
implies the stability of the totally discretized method for most
standard methods of lines~\cite{Kreiss-Wu}, e.g. with the use
of a Runge-Kutta time integrator. 

\subsection{The Cauchy Problem}

We consider the Cauchy problem for a second order system with constant
(possibly complex) coefficients in $s$ space dimensions,
\begin{equation}\label{system-2D}
    {\bf u}_{tt}=\sum_{j,k=1}^s 
   A_{jk} \frac{\partial}{\partial {\tilde x}_j}
   \frac{\partial}{\partial {\tilde x}_k} {\bf u} 
     := P_0(\partial/\partial {\tilde x}){\bf u},
     \quad {\tilde{\bf x}}=({\tilde x}_1,\dotsc,{\tilde x}_s )
     \in \mathbb{R}^s, \quad t\ge0,
\end{equation}
with the initial conditions
\begin{equation}
   {\bf u}({\tilde{\bf x}},t=0)={\bf f}({\tilde{\bf x}}),\quad {\bf u}_t({\tilde{\bf x}},t=0)
   ={\bf g}({\tilde{\bf x}}), \quad {\bf u}, {\bf f}, {\bf g} \in \mathbb{C}^n.
\end{equation}
(We abbreviate $\partial_\alpha u =u_\alpha$ where confusion does not
arise.) Here, for each $(j,k)$, $A_{jk}$ are constant Hermitian
matrices $ \in \mathbb{C}^{n,n}$, and the data ${\bf f}={\bf
f}({\tilde{\bf x}})$ and ${\bf g}={\bf g}({\tilde{\bf x}})$ are smooth
and $1$-periodic in each ${\tilde x}_j$, $j=1,\dotsc,s$. The solution
${\bf u}={\bf u}({\tilde{\bf x}},t)$ is then smooth and $1$-periodic
in each ${\tilde x}_j$. Moreover, we consider solutions with
$\int_{\mathbb{R}^s}{\bf u} \, d{\tilde{\bf x}}=0$.

We assume that the Hermitian operator $P_0$ in (\ref{system-2D}) is
elliptic, i. e. there exists a positive constant $\delta$ such that
\begin{equation}\label{ellipticity}
     \sum_{j,k=1}^s A_{jk} \xi_j \xi_k \ge \delta |\xi|^2 I
\end{equation}
for all vectors $\xi \in \mathbb{R}^s$. Here $I$ is the $n\times n$
identity matrix.

We introduce a shift by 
$$
  \tilde{\bf{x}}={\bf{x}}+{{\bar \beta}} \, t,\quad {\bf{x}}= 
 (x_1,\dotsc,x_s )\in \mathbb{R}^s ,\quad {{\bar \beta}}= 
   (\beta^1,\dotsc,\beta^s)\in \mathbb{R}^s,\quad \beta^j>0,
$$
and obtain the shifted system
\begin{equation}\label{shifted-system-2D}
  {\bf u}_{tt}=2P_1(\partial/\partial x){\bf u}_t
    -P_1^2(\partial/\partial x){\bf u}
  +P_0(\partial/\partial x){\bf u}.
\end{equation}
Here $P_1$ is a scalar operator,  
\begin{equation*}\label{P1}
         P_1(\partial/\partial x)
           =\sum_{j=1}^s \beta^j \frac{\partial}{\partial x_j}.
\end{equation*}
\\
\\
\noindent
{\bf Theorem 1.} The Cauchy problem for (\ref{shifted-system-2D}) is well-posed.

\noindent
{\it Proof.} If we set ${\bf v}={\bf u}_t-P_1(\partial/\partial x){\bf
u}$, we get the first order system
\begin{equation}\label{1st-order-system-2D}
\left( \begin{array}{c}
{\bf u}\\
{\bf v} 
\end{array} \right)_t= P_1(\partial/\partial x)\left( \begin{array}{c}
{\bf u}\\
{\bf v} 
\end{array} \right)+\left( \begin{array}{c c}
0 & I\\
P_0(\partial/\partial x) & 0
\end{array} \right)\left( \begin{array}{c}
{\bf u}\\
{\bf v} 
\end{array} \right).
\end{equation}

We Fourier transform (\ref{1st-order-system-2D}) and get
\begin{equation}\label{1st-order-system-2D_Fourier}
\left( \begin{array}{c}
{\bf \hat{u}}\\
{\bf \hat{v}} 
\end{array} \right)_t= \hat{P_1}(i\omega)\left( \begin{array}{c}
{\bf \hat{u}}\\
{\bf \hat{v}} 
\end{array} \right)+\left( \begin{array}{c c}
0 & I\\
-\hat{P_0}(\omega) & 0
\end{array} \right)\left( \begin{array}{c}
{\bf \hat{u}}\\
{\bf \hat{v}} 
\end{array} \right), \quad \omega\neq {\bf 0},
\end{equation}
where $\omega=(\omega_1,\dotsc,\omega_s)\in \mathbb{R}^s$ and
$\hat{P_1}(i\omega)=i\sum_{j=1}^s \beta^j \omega_j$ and $\hat{P_0}(\omega)=\sum_{j,k=1}^s
A_{jk} \omega_j \omega_k$. Since $P_0$ is elliptic,
we have $\hat{P_0}=\hat{P_0}^* \ge \delta_0 |\omega|^2 I$, for some $\delta_0>0$.
We can then introduce new variables
\begin{equation}\label{T}
{\bf \hat{w}}=T \left( \begin{array}{c}
{\bf \hat{u}}\\
{\bf \hat{v}} 
\end{array} \right), \quad T= \left( \begin{array}{c c}
I & 0\\
0 & \hat{P_0}^{-1/2}
\end{array} \right)
\end{equation}
and, since $T$ and $\hat{P_1}$ commute, we obtain from (\ref{1st-order-system-2D_Fourier}) 
\begin{equation}\label{1st-order-system}
   {\bf \hat{w}}_t= \hat{P_1}(i\omega){\bf \hat{w}}+\left(
    \begin{array}{c c} 0 & \hat{P_0}^{1/2}(\omega)\\
        -\hat{P_0}^{1/2}(\omega) & 0
     \end{array} \right){\bf \hat{w}}:=S {\bf \hat{w}}.
\end{equation}

Since the matrix $S$ is skew Hermitian, $S^*=-S$, we obtain
\begin{equation*}
   \frac{\partial}{\partial t}||{\bf \hat{w}}||^2
    =(S {\bf \hat{w}},{\bf \hat{w}})
     +({\bf \hat{w}},S{\bf \hat{w}})
     =({\bf \hat{w}},(S^*+S){\bf \hat{w}})=0.
\end{equation*}

Therefore, by Parseval's relation, there is an energy estimate
and the Cauchy problem is well-posed~\cite{green-book}. $\Box$
\\

Now we show how to construct stable finite difference
approximations to (\ref{shifted-system-2D}). We leave time
continuous and use the method of lines. For brevity, we treat
the case $\beta^j > 0$.

Let $h_j=1/N_j$, $j=1,\dotsc,s$, denote spatial gridlengths,
where $N_j$ are natural numbers. For any multi-index
$\nu=(\nu_1,\dotsc,\nu_s) \in \mathbb{Z}^s$, let ${\bf
x}_{\nu}=(h_1 \nu_1,\dotsc,h_s \nu_s)$ denote the corresponding
gridpoint. We consider gridfunctions ${\bf u}_{\nu}:={\bf
u}_{\nu}({\bf x}_{\nu},t)$ approximating ${\bf u}({\bf
x}_{\nu},t)$ and introduce a translation operator $E_j$ in the
$j$-th coordinate by 
$$
  E_j^p {\bf u}_{\nu}
   ={\bf u}_{\nu}({\bf x}_{\nu}+ph_j{\bf e}_j,t),
  \quad p \in  \mathbb{Z},
$$ 
where ${\bf e}_j=(0,\dotsc,0,1,0,\dotsc,0)$ is the vector
containing a $1$ in the $j$-th position and zeros elsewhere. We
then define the forward, backward, and the central difference
operators in the $j$-th coordinate direction by
$$
  h_j D_{+j}=E_j^1 - E_j^0, \quad h_j D_{-j}=E_j^0 - E_j^{-1},
     \quad 2D_{0j} = D_{+j} + D_{-j}.
$$

We approximate (\ref{shifted-system-2D}) by
\begin{equation}\label{approximation1-nonelliptic}
      {\bf u}_{\nu tt}=2p_1(D){\bf u}_{\nu t}-p_1^2(D)
      {\bf u}_{\nu} +p_0(D){\bf u}_{\nu},
\end{equation}
where $p_0(D)$ is the centered approximation
\begin{equation}\label{P0_approx}
     p_0(D)=\sum_{j=1}^s A_{jj} D_{+j} D_{-j}
   +\sum_{j \neq k=1}^s A_{jk} D_{0j} D_{0k},
\end{equation}
and $p_1(D)$ is any one of the following approximations:\\
1) Centered approximation,
\begin{equation}\label{P1_approx1}
p_1(D)=\sum_{j=1}^s \beta^j D_{0j},
\end{equation}
2) First order accurate one-sided approximation, 
\begin{equation}\label{P1_approx2}
         p_1(D)=\sum_{j=1}^s \beta^j D_{+j},
\end{equation}
3) Second order accurate one-sided approximation, 
\begin{equation}\label{P1_approx3}
           p_1(D)=\sum_{j=1}^s \beta^j  D_{pj},
\end{equation} 
where
\begin{equation}
   D_{pj}=D_{+j} -\frac{h_j}{2}D_{+j}^2.
   \label{eq:dpj}
\end{equation}

\noindent
{\bf Remark.} It is not necessary to assume that $\beta^j>0$ in
(\ref{P1_approx1})-(\ref{P1_approx3}). In general, we can use
$\frac{\beta^j+|\beta^j|}{2}D_{+j} + \frac{\beta^j-|\beta^j|}{2}D_{-j}$
in (\ref{P1_approx2}). For the second order one-sided approximation
(\ref{P1_approx3}), we replace $D_{+j}$ and $D_{-j}$ by $D_{pj}$ and
$D_{mj}=D_{-j}+\frac{h_j}{2}D_{-j}^2$, respectively. 
\\
\\
\noindent
{\bf Theorem 2.} The approximation (\ref{approximation1-nonelliptic}) is stable.

\noindent
{\it Proof.} As in the continuum case, we write
(\ref{approximation1-nonelliptic})
as a first order system and Fourier transform to get 
\begin{equation}\label{approx-1st-order-system-Fourier}
\left( \begin{array}{c}
{\bf \hat{u}}\\
{\bf \hat{v}} 
\end{array} \right)_t= \hat{p}_1\left( \begin{array}{c}
{\bf \hat{u}}\\
{\bf \hat{v}} 
\end{array} \right)+\left( \begin{array}{c c}
0 & I\\
-\hat{p}_0 & 0
\end{array} \right)\left( \begin{array}{c}
{\bf \hat{u}}\\
{\bf \hat{v}} 
\end{array} \right),
\end{equation}
where 
\begin{equation}\label{P0_hat}
  \hat {p}_0=\sum_{j=1}^s A_{jj}\frac{4}{h_j^2} \sin^2\frac{\xi_j}{2}
  +\sum_{j \neq k=1}^s A_{jk}\frac{1}{h_jh_k} \sin \xi_j \sin \xi_k, \quad
   \xi_j=\omega_j h_j,\quad |\xi_j|\le \pi,
\end{equation}
and $\hat{p}_1$ is one of the following:
\begin{equation}\label{P1_hat1}
\hat{p}_1=\sum_{j=1}^s \beta^j \frac{1}{h_j} i \sin \xi_j,
\end{equation}
\begin{equation}\label{P1_hat2}
\hat{p}_1=\sum_{j=1}^s \beta^j \frac{1}{h_j} \left(i \sin \xi_j
-2\sin^2\frac{\xi_j}{2}\right),
\end{equation}
\begin{equation}\label{P1_hat3}
 \hat{p}_1=\sum_{j=1}^s \beta^j \frac{1}{h_j} 
  \left(i \sin \xi_j(1-2\sin^2\frac{\xi_j}{2})-4\sin^4\frac{\xi_j}{2}\right),
\end{equation} 
corresponding to (\ref{P1_approx1})-(\ref{P1_approx3}).

Since
\begin{equation}
  \sin^2\xi= 4\sin^2{\xi\over 2} \cos^2{\xi\over 2} 
     =4 \sin^2 {\xi\over 2} -4 \sin^4 {\xi\over 2},
\end{equation}
we have
\begin{equation}\label{phat0}
\hat {p}_0=\sum_{j,k=1}^s A_{jk}\frac{1}{h_jh_k} \sin \xi_j \sin \xi_k+\sum_{j=1}^s
     A_{jj}\frac{4}{h_j^2} \sin^4\frac{\xi_j}{2}.
\end{equation}
From the ellipticity condition (\ref{ellipticity}) it follows that
$\hat{p}_0$ is positive. As $\xi_j \rightarrow 0$ we have $\sin
{\xi_j}/h_j \rightarrow \omega_j$. Therefore, the first sum in
(\ref{phat0}) is strictly positive. When $|\xi_j| = \pi$, the first sum
in (\ref{phat0}) is zero but the second sum is not because
$\sin{\frac{\xi_j}{2}}\neq 0$. Therefore $\hat{p}_0$ is positive
definite, and we can use the same transformation as in (\ref{T}) and
write (\ref{approx-1st-order-system-Fourier}) as
\begin{equation}\label{approx-1st-order-system}
    {\bf \hat{w}}_t= \hat{p}_1{\bf \hat{w}}
    +\left( \begin{array}{c c}
   0 & \hat{p}_0^{1/2}(\omega)\\
    -\hat{p}_0^{1/2}(\omega) & 0
\end{array} \right){\bf \hat{w}}.
\end{equation}

The second term on the right hand side of (\ref{approx-1st-order-system})
is again skew Hermitian and has no influence on the stability. Thus, we
need only consider 
$$
{\bf \hat{w}}_t=\hat{p}_1{\bf \hat{w}},
$$
which consists of difference approximations of scalar equations of the
above type. To show that the approximations
(\ref{P1_approx1})-(\ref{P1_approx3}) are stable, we set
$\hat u=e^{\lambda t} \hat u_0$ and get $\lambda=\hat{p}_1$.
By (\ref{P1_hat1})-(\ref{P1_hat3}), we have $\Re \lambda \le 0$ and there are
no exponentially growing modes. $\Box$
\\

The approximation (\ref{approximation1-nonelliptic}) involves a wide stencil.
Therefore extra boundary conditions (ghost points) are required and the
resulting accuracy is less than with a more compact stencil. In order to
investigate other approximations with a more compact stencil, we write
(\ref{shifted-system-2D}) as 
\begin{equation}\label{shifted-system-2D-P}
   {\bf u}_{tt}=2P_1(\partial/\partial x){\bf u}_t
  +P(\partial/\partial x){\bf u},\quad P(\partial/\partial x)
    =P_0(\partial/\partial x)-P_1^2(\partial/\partial x)
\end{equation}
and approximate it by
\begin{equation}\label{approximation1-elliptic}
{\bf u}_{\nu tt}=2p_1(D){\bf u}_{\nu t}+p(D){\bf u}_{\nu},
\end{equation}
where $p_1(D)$ is given by (\ref{P1_approx1}) and $p(D)$ is the centered approximation
\begin{equation}\label{P_approx1}
 p(D)=\sum_{j=1}^s \left( A_{jj}-{\beta^j}^2 \right) D_{+j} D_{-j}
 +\sum_{j \neq k=1}^s \left(A_{jk}-\beta^j \beta^k\right) D_{0j} D_{0k}.
\end{equation}
\\
\\
\noindent
{\bf Theorem 3.} The approximation (\ref{approximation1-elliptic}) is
stable if $A_{jj}-{\beta^j}^2 >0$.

\noindent
{\it Proof.} We write (\ref{approximation1-elliptic}) as
\begin{equation}\label{approximation1-elliptic2}
   {\bf u}_{\nu tt}=2p_1(D){\bf u}_{\nu t}-p_1^2(D){\bf u}_{\nu}+q(D){\bf u}_{\nu},
   \quad q(D)=p(D)+p_1^2(D).
\end{equation}
We use the relation $D_{+j} D_{-j}=D_{0j}^2-\frac{h_j^2}{4}D^2_{+j} D^2_{-j}$
 and write 
\begin{align*}\label{q_approx}
   q(D)&=\sum_{j,k=1}^s A_{jk} D_{0j} D_{0k}-\frac{1}{4}\sum_{j=1}^s
  (A_{jj}-{\beta^j}^2)h_j^2 D^2_{+j} D^2_{-j}.
\end{align*}
In the same way as in the continuum case, we
write (\ref{approximation1-elliptic2})
as a first order system and Fourier transform to get 
\begin{equation}\label{approx-1st-order-system-Fourier-1}
\left( \begin{array}{c}
{\bf \hat{u}}\\
{\bf \hat{v}} 
\end{array} \right)_t= \hat{p}_1\left( \begin{array}{c}
{\bf \hat{u}}\\
{\bf \hat{v}} 
\end{array} \right)+\left( \begin{array}{c c}
0 & I\\
-\hat{q} & 0
\end{array} \right)\left( \begin{array}{c}
{\bf \hat{u}}\\
{\bf \hat{v}} 
\end{array} \right),
\end{equation}
where 
\begin{equation}\label{q_hat}
  \hat {q}=\sum_{j,k=1}^s A_{jk}\frac{1}{h_j h_k} \sin\xi_j \sin\xi_k+\sum_{j=1}^s
  \left(A_{jj}-{\beta^j}^2\right)\frac{4}{h_j^2} \sin^4\frac{\xi_j}{2}, \quad \xi_j
  =\omega_j h_j,\quad |\xi_j|\le \pi
\end{equation}
and $\hat{p}_1$ is given by (\ref{P1_hat1}). By the ellipticity
condition (\ref{ellipticity}), it is clear that $\hat{q}$ is a
positive definite matrix if $A_{jj}-{\beta^j}^2 >0$. Therefore,
we can use the same transformation as in (\ref{T})
and write (\ref{approx-1st-order-system-Fourier-1}) as
\begin{equation}\label{approx-1st-order-system-1}
   {\bf \hat{w}}_t= \hat{p}_1{\bf \hat{w}}+\left( \begin{array}{c c}
    0 & \hat{q}^{1/2}(\omega)\\
       -\hat{q}^{1/2}(\omega) & 0
       \end{array} \right){\bf \hat{w}}.
\end{equation}

The second term on the right hand side of
(\ref{approx-1st-order-system-1}) is again skew Hermitian and has no
influence on the stability. Thus, we need only to consider 
$$
{\bf \hat{w}}_t=\hat{p}_1{\bf \hat{w}},
$$
and the stability follows in the same way as in Theorem 2. $\Box$
\\

\noindent 
{\bf Remark.} If the operator $P$ is elliptic, we have
$A_{jj}-{\beta^j}^2 >0$, and by Theorem 3 the approximation
(\ref{approximation1-elliptic}) is stable. However, it is possible to
have $A_{jj}-{\beta^j}^2 >0$ while $P$ is non-elliptic. In this case,
the approximation (\ref{approximation1-elliptic}) remains stable when
$P$ is non-elliptic. In other words, the stability of
(\ref{approximation1-elliptic}) does not depend upon the coefficients of
mixed derivatives $A_{jk},\, j \ne k$.
\\

\noindent
{\bf Remark.} In the scalar case, (\ref{approximation1-elliptic}) reduces
to the $W$-algorithm (\ref{eq:subwave}).
\\

In the excision problem, we use the subluminal algorithm
(\ref{approximation1-elliptic}) in the subluminal region where
$A_{jj}-{\beta^j}^2 >0$. In the superluminal region where the shift $\beta^j$
is large so that $A_{jj}-{\beta^j}^2 \le 0$, we use the superluminal algorithm
(\ref{approximation1-nonelliptic}) instead. We need then a prescription for
switching from one algorithm to the other. There are two distinct ways to do
this. One is to make a sharp switch between the algorithms where the transition
from superluminal to subluminal region takes place. The other, used
in~\cite{excis}, is to introduce a smooth, monotonic blending function and use
a blended algorithm, which turns into the superluminal algorithm inside the
superluminal region and reduces monotonically to the subluminal algorithm in
the outside. For this purpose, note that the superluminal algorithm remains
stable in the subluminal region.   

As a further alternative to the above approximations, we can approximate
(\ref{shifted-system-2D-P}) by adding a fourth differential order term
\begin{equation}\label{approximation2} 
    {\bf u}_{\nu tt}=2p_1(D){\bf u}_{\nu t}
    +p(D){\bf u}_{\nu}-Q(D){\bf u}_{\nu}, 
\end{equation} 
where $p_1(D)$ is given by (\ref{P1_approx1}) and
\begin{equation}\label{Q_approx}
    Q(D)=\frac{1}{4}\sum_{j=1}^s \alpha_j h_j^2 D^2_{+j} D^2_{-j}, 
   \quad  \alpha_j\ge0. 
\end{equation}
The motivation for adding such a fourth order term is to modify the matrix
$\hat{q}$ in (\ref{q_hat}) so that it becomes positive definite even if
$A_{jj}-{\beta^j}^2 \le 0$. When $A_{jj}-{\beta^j}^2 >0$, the matrix $\hat{q}$
is positive definite and this added term is unnecessary. We can take advantage
of this by embedding the switch or blending function in the choice of
$\alpha_j$, with $\alpha_j=0$ in the outer region. 
\\
\\
\noindent
{\bf Theorem 4.} The approximation (\ref{approximation2}) is stable if
$A_{jj}+\alpha_j I \ge {\beta^j}^2 I$.

\noindent
{\it Proof.} We use the relation $D_{0j}^2=D_{+j} D_{-j}+\frac{h_j^2}{4}D^2_{+j} D^2_{-j}$
and write 
\begin{align*}\label{P_approx-2}
    p(D)&=\sum_{j,k=1}^s (A_{jk}-\beta^j \beta^k) D_{0j}
    D_{0k}-\frac{1}{4}\sum_{j=1}^s
    (A_{jj}-{\beta^j}^2)h_j^2 D^2_{+j} D^2_{-j}\\
   &=-p_1^2(D)+\sum_{j,k=1}^s A_{jk} D_{0j}
     D_{0k}-\frac{1}{4}\sum_{j=1}^s
   (A_{jj}-{\beta^j}^2)h_j^2 D^2_{+j} D^2_{-j}.
\end{align*}

We can then write (\ref{approximation2}) as
\begin{equation}\label{approximation2-2}
   {\bf u}_{\nu tt}=2p_1(D){\bf u}_{\nu t}
     -p_1^2(D){\bf u}_{\nu}+q(D){\bf u}_{\nu},
\end{equation}
where 
\begin{equation}\label{q_approx}
   q(D)=\sum_{j,k=1}^s
     A_{jk} D_{0j} D_{0k}-\frac{1}{4}\sum_{j=1}^s
    (A_{jj}-{\beta^j}^2+\alpha_j) h_j^2 D^2_{+j} D^2_{-j}.
\end{equation}

In the same way as before, we write (\ref{approximation2-2}) as
a first order system and Fourier transform to get 
\begin{equation}\label{1st-order-system-2D_Fourier-approx2}
\left( \begin{array}{c}
{\bf \hat{u}}\\
{\bf \hat{v}} 
\end{array} \right)_t= \hat{p}_1\left( \begin{array}{c}
{\bf \hat{u}}\\
{\bf \hat{v}} 
\end{array} \right)+\left( \begin{array}{c c}
0 & I\\
-\hat{q} & 0
\end{array} \right)\left( \begin{array}{c}
{\bf \hat{u}}\\
{\bf \hat{v}} 
\end{array} \right),
\end{equation}
where
\begin{equation*}
\hat{q}=\sum_{j,k=1}^s A_{jk}\frac{1}{h_jh_k} \sin \xi_j 
     \sin \xi_k+\sum_{j=1}^s
  (A_{jj}-{\beta^j}^2+\alpha_j)\frac{4}{h_j^2}
    \sin^4\frac{\xi_j}{2}.
\end{equation*}
If $A_{jj}+\alpha_j I \ge {\beta^j}^2 I$ then, because of
ellipticity, $\hat {q}$ is positive definite and stability
follows in the same way as before. $\Box$


\subsection{Half-plane Problems}

We consider the scalar wave equation with constant coefficients
in two space dimensions,
\begin{equation}\label{wave-2D}
u_{\tilde{t}\tilde{t}}=a_1\,u_{\tilde{x}\tilde{x}}
+2b_1\,u_{\tilde{x}\tilde{y}}
  +c_1\,u_{\tilde{y}\tilde{y}}:=P_0 u.
\end{equation}
In the moving coordinate system, $t=\tilde{t}$,
$x=\tilde{x}-\beta^x\tilde{t}$, $y=\tilde{y}-\beta^y\tilde{t}$, with
$\beta^x,\beta^y>0$, we get the shifted wave equation,
\begin{equation}\label{2D-3}
    u_{tt}=2(\beta^xu_{xt}+\beta^y u_{yt})+au_{xx}
    +2bu_{xy}+cu_{yy}:= 2 P_1 u_t + P u.
\end{equation} Here the coefficients $a=a_1-{\beta^x}^2$,
$b=b_1-\beta^x\beta^y$, and $c=c_1-{\beta^x}^2$ are assumed to
be constant. Moreover, we assume that the space operator $P_0$
in (\ref{wave-2D}) is elliptic, namely $a_1>0$ and $c_1>0$ and
$b_1^2<a_1 c_1$. Therefore, by Theorem 1, the Cauchy problem for
(\ref{2D-3}) is well-posed.

We consider (\ref{2D-3}) in the half-space
$$
    0\le x<\infty, \quad -\infty< y <\infty, \quad t\ge 0
$$
and we assume that $u$ is $1$-periodic in $y$. The number of
boundary conditions needed at $x=0$ is equal to the number of
outgoing characteristics of the equation
$u_{tt}=2\beta^xu_{xt}+au_{xx}$. We consider two distinct
half-plane problems determined by the coefficients of the
operator $P$.
\\

\noindent
{\bf Half-plane problem I}: If $a>0$ and $b^2<ac$, then the
operator $P$ is elliptic and one boundary condition is needed
at $x=0$. In the excision problem, this is the case of
subluminal shift with a timelike boundary.
\\
\\
\noindent
{\bf Half-plane problem II}: If $a<0$, then the operator $P$ is
non-elliptic. In the excision problem, this is the case of a
superluminal shift with a spacelike boundary.

\subsubsection{Half-plane Problem I (Subluminal Case)}

This is the problem treated in~\cite{harm} by the energy method.
In the present context of (\ref{2D-3}), the energy is given by
\begin{equation}
     E=\|u_t\|^2+a\|u_x\|^2+2b(u_x,u_y)+c\|u_y\|^2
\end{equation}
in terms of the $L_2$ scalar product and the corresponding
norm
\begin{equation}\label{L2-product}
   (v,w)=\int_{0}^{1}\int_0^{\infty}v\,w\,dx \, dy
       ,\quad\|v\|^2=(v,v).
\end{equation} 
If $u$ solves (\ref{2D-3}), then integration by parts gives
\begin{equation}
   \partial_t E=-2u_t(\beta^xu_t+au_x+bu_y)\Bigl\vert_{x=0}.
\label{eq:econs}
\end{equation}
Any boundary condition satisfying the dissipative condition
$\partial_t E \le 0$ gives an energy estimate sufficient to
establish the well-posedness of the Cauchy problem, including
the Dirichlet condition
\begin{equation}\label{bc1_2D}
       u_t(0,y,t)=0
\end{equation}
     and the Neumann condition
\begin{equation}\label{bc2_2D}
     \beta^xu_t(0,y,t)+au_x(0,y,t)+bu_y(0,y,t)=0
\end{equation}
for which energy is conserved.

As difference approximation for the half-plane problem, we use
(\ref{approximation1-elliptic}), which in the present case reduces to
the $W$-algorithm (\ref{eq:subwave}). By introducing a discrete energy
norm and using summation by parts, a discrete version of
(\ref{eq:econs}) has been used to establish stability of the finite
difference problem. For details we refer to~\cite{harm}. 

\subsubsection{Half-plane Problem II (Superluminal Case)}

To investigate the well-posedness of the continuum problem, we
use mode analysis. We apply a Laplace transformation in $t$ and
Fourier transformation in $y$.  \\
\\
\noindent
{\bf Theorem 5.} The half-plane problem (\ref{2D-3}) with $a<0$ is well-posed. 

\noindent
{\it Proof.} By substituting $u=\hat u(x) \, e^{st+i\omega y}$, $s \in \mathbb{C}$,
$\omega \in \mathbb{R}$, into (\ref{2D-3}) we obtain
\begin{equation}\label{ode}
    a \hat u_{xx}+(2ib\omega+2\beta^xs)\hat u_x
     +(2i\beta^y\omega s-s^2 -c\omega^2)\hat u=0.
\end{equation}
The general solution to the ordinary differential equation
(\ref{ode}) is of the form $\hat u(x)=\sigma_1 e^{\kappa_1
x}+\sigma_2 e^{\kappa_2 x}$, where $\kappa_1$ and $\kappa_2$ are
the solutions of the characteristic equation
\begin{equation}\label{char-pde}
  a \kappa^2+(2bi\omega +2\beta^xs)\kappa+2i\beta^y\omega s-s^2-c\omega^2=0.
\end{equation}

Without restriction we can assume $a=-1$. Moreover, since the
sign of $\Re \kappa$ does not depend on $\omega$, we set $\omega
=0$. We then obtain
\begin{equation*}
 \kappa_{1,2}=\beta^x s \pm \sqrt{({\beta^x}^2-1)s^2}.
\end{equation*}
For $\Re s>0$, we have $\Re \kappa_{1,2}>0$ and there is no
bounded solution $\hat u$. Therefore no boundary condition is
needed and the problem is well-posed. $\Box$ \\

As difference approximation for the half-plane problem, we
can use either (\ref{approximation1-nonelliptic}) or
(\ref{approximation2}). We study the stability of the
approximations by mode analysis. Below we show that
(\ref{approximation1-nonelliptic}) is stable with $p_1(D)$ in
(\ref{P1_approx2}). The stability of the other approximations with
$p_1(D)$ in (\ref{P1_approx1}) and (\ref{P1_approx3}) can be
shown in the same way. 

On a uniform spatial grid $\Omega_h=(\nu h,\mu h), \,
\nu=0,1,2,\dotsc, \, \mu=1,2,\dotsc,N$, with spacing $h$, let
$v(t):=u_{\nu \mu}(t)$ be the gridfunction approximating
$u(x_{\nu},y_{\mu},t)$. We consider the shifted wave equation
(\ref{2D-3}) and approximate it by 
\begin{equation}\label{2D-3-left-approximate}
  v_{tt}=2(\beta^x D_{+x}+\beta^y D_{+y})v_{t}-(\beta^xD_{+x}
   +\beta^yD_{+y})^2v+(a_1 D_{+x}D_{-x}
        +2b_1 D_{0x}D_{0y}+c_1 D_{+y}D_{-y})v,
\end{equation}
for $\nu=1,2,\dotsc$. For every fixed $\mu$, we need one extra
boundary condition to determine $u_{0\mu}$. We use a third order
extrapolation
\begin{equation}\label{extrapolation}
      h^3 D^3_{+x}u_{0\mu}=0.
\end{equation}

We consider bounded solutions of type
\begin{equation}\label{ansatz}
 u_{\nu\mu}(t)=e^{st+i\omega \mu h}\varphi_\nu, 
     \quad ||\varphi||_h<\infty.
\end{equation}
Putting (\ref{ansatz}) into (\ref{2D-3-left-approximate}), we
get the eigenvalue problem
\begin{multline}\label{eig} \varphi_\nu
     s^2-2\frac{\beta^x}{h}(\varphi_{\nu+1}-\varphi_{\nu}) s
 -2\frac{\beta^y}{h}(i\sin{\xi}
    -2\sin^2{\frac{\xi}{2}})\varphi_\nu s\\
  +\frac{{\beta^x}^2}{h^2}(\varphi_{\nu+2}
   -2\varphi_{\nu+1}+\varphi_{\nu})
  +2\frac{\beta^x\beta^y}{h^2}(\varphi_{\nu+1}-\varphi_{\nu})
    (i\sin{\xi}-2\sin^2{\frac{\xi}{2}})+\frac{{\beta^y}^2}{h^2}
   (i\sin^2{\xi} -2\sin^2{\frac{\xi}{2}})^2\varphi_\nu\\
   -\frac{a_1}{h^2}(\varphi_{\nu+1}
     -2\varphi_\nu+\varphi_{\nu-1})
  -\frac{b_1}{h^2} i\sin{\xi}(\varphi_{\nu+1}-\varphi_{\nu-1})
    +4\frac{c_1}{h^2} \sin^2{\frac{\xi}{2}}\varphi_\nu=0,
    \quad \xi=\omega h.
\end{multline}

The approximation (\ref{2D-3-left-approximate})-(\ref{extrapolation})
is stable if and only if the Kreiss condition is satisfied, or
equivalently if (\ref{eig}) has no eigenvalue $s$ with $\Re s \ge
0$~\cite{brown-book}.  The constant-coefficient ordinary difference
equation (\ref{eig}) has solution of the form
$$
   \varphi_{\nu}=\sum_{j=1}^{3}\sigma_j \kappa_j^{\nu},
$$
where $\kappa_j$ are the three solutions of the characteristic equation
\begin{eqnarray}\label{char}
        s^2 &-& 2\left(\frac{\beta^x}{h}(\kappa - 1)
    +\frac{\beta^y}{h}(i\sin{\xi}
    -2\sin^2{\frac{\xi}{2}})\right)s
	+\left(\frac{\beta^x}{h}(\kappa - 1)
    +\frac{\beta^y}{h}(i\sin{\xi}
    -2\sin^2{\frac{\xi}{2}})\right)^2  \nonumber \\
     &-& \frac{a_1}{h^2}\frac{(\kappa-1)^2}{\kappa}
     -\frac{b_1}{h^2} (\kappa-\frac{1}{\kappa})i\sin{\xi}
       +4\frac{c_1}{h^2} \sin^2{\frac{\xi}{2}}=0.
\end{eqnarray}

By Lemma 12.1.6 of~\cite{brown-book}, for $\Re s>0$ the
characteristic equation (\ref{char}) has no solutions with
$|\kappa|=1$ and there is exactly one solution with
$|\kappa|<1$. Roughly speaking, the number of left points in the
difference stencil determines the number of solutions to the
characteristic equation with $|\kappa|<1$. We call this solution
$\kappa_1$ and write the bounded solution as
\begin{equation}\label{bounded-sol}
   u_{\nu}(t)=e^{st+i\omega \mu h}\sigma_1\kappa_1^{\nu}.
\end{equation}      
By substituting (\ref{bounded-sol}) into the boundary condition
(\ref{extrapolation}), we get
\begin{equation}\label{BC-kappa-2D}
     \sigma_1 (\kappa_1-1)^3 e^{st+i\omega \mu h}=0.
\end{equation}
Since $\kappa_1\neq 1$ for $\Re s>0$, (\ref{BC-kappa-2D}) has
only the trivial solution $\sigma_1=0$. Now, we let $\kappa
\rightarrow 1$ and investigate if there is any sequence $\{s\}$
such that $\Re s \rightarrow 0$ with $\Re s>0$. We then get from
(\ref{char})
\begin{equation}
   {\tilde s}^2-2\beta^y(i\sin{\xi}
      -2\sin^2{\frac{\xi}{2}}){\tilde s}
      +{\beta^y}^2(i\sin{\xi}-2\sin^2{\frac{\xi}{2}})^2
      +4 c_1 \sin^2{\frac{\xi}{2}}=0, \quad {\tilde s} =sh,
\end{equation}
and therefore
\begin{equation}
    {\tilde s}=\beta^y(i\sin{\xi}-2\sin^2{\frac{\xi}{2}})
    \pm \sqrt{-4c_1 \sin^2{\frac{\xi}{2}}}.
\end{equation}
Since $\beta^y>0$ and $c_1>0$, we have $\Re s<0$ if $\xi \nrightarrow 0$.
In the case where $\xi \rightarrow 0$, we get from (\ref{char})
\begin{equation}\label{1D-char}
    s^2-2\frac{\beta^x}{h} s (\kappa -1)
      +\frac{{\beta^x}^2}{h^2}(\kappa - 1)^2
      -\frac{a_1}{h^2}\frac{(\kappa-1)^2}{\kappa}=0.
\end{equation}
Letting $s\rightarrow 0$, we then get from (\ref{1D-char}) that
$\kappa_{1,2}=1$ and $\kappa_3=a_1/{\beta^x}^2<1$. Since
for $\Re s>0$ there is no solution with $|\kappa|=1$, the only
solution is $\kappa_3$ which is strictly less than 1 and does
not converge to 1. Therefore there is no positive sequence $\{s\}$
such that $\Re s \rightarrow 0$ for $|\xi|\le\pi$. Now, we can
prove the following theorem:
\\
\\
\noindent
{\bf Theorem 6.} The approximation
(\ref{2D-3-left-approximate})-(\ref{extrapolation}) is stable. 

\noindent {\it Proof.} Since there is no eigenvalue $s$ with
$\Re s \ge 0$ to the eigenvalue problem (\ref{eig}) giving
bounded solutions (\ref{ansatz}), the Kreiss condition is
satisfied and stability follows. $\Box$


\section{Tests of the superluminal algorithms}
\label{sec:periodic}

In the subluminal case where  the evolution proceeds in a timelike
direction, the $W$-algorithm (\ref{eq:subwave}) provides an
accurate, flux-conservative, second order treatment of the IBVP.
This was proved for a 1D quasilinear wave equation in~\cite{excis}
using the discrete energy method. In~\cite{babev,harm}, the results
were extended to the 3D case and applied to the harmonic Einstein
system (\ref{eq:nlg}). The semi-discrete conservation laws extend to
the principle part of the harmonic Einstein system and contribute to
excellent long term performance in test problems. We use this
$W$-algorithm to treat the outer region of the model excision
problem considered in Sec.~\ref{sec:excision}. 

In this model problem, the inner boundary is chosen to be spacelike,
corresponding to the strategy for excising an interior singularity.
The evolution near the inner boundary proceeds in a spacelike
direction (superluminal shift) so that the spatial grid tracks the
boundary. For this superluminal case, the $W$-algorithm is unstable
and one of the algorithms considered in Sec.~\ref{sec:algorithms}
must be used.  These algorithms are either given by
(\ref{approximation1-nonelliptic}), with $p_1(D)$ given by one of
the  approximations (\ref{P1_approx1})-(\ref{P1_approx3}), or by
(\ref{approximation2}).

In the case of the 2D shifted wave equation (\ref{eq:2wave}),
the choice (\ref{P1_approx1}) reduces to the centered algorithm
\begin{equation}
    V:= \bigg ((\partial_t -\beta^x D_{0x} -\beta^yD_{0y} )^2
         -a_1D_{+x} D_{-x}-c_1D_{+y} D_{-y} -2b_1 D_{0x} D_{0y}
	 \bigg)u   =0 ;
      \label{eq:vwave}
\end{equation}
the choice (\ref{P1_approx2}) reduces to
\begin{equation}
    V_+:= \bigg ((\partial_t -\beta^x D_{+x} -\beta^yD_{+y} )^2
         -a_1D_{+x} D _{-x}-c_1D_{+y} D_{-y} -2b_1 D_{0x} D_{0y}
        \bigg)u   =0 ,
      \label{eq:v+wave}
\end{equation}
in which the shift terms are treated by first order accurate
one-sided difference  operators; the choice (\ref{P1_approx3})
reduces to 
\begin{equation}
    V_p:= \bigg ((\partial_t -\beta^x D_{px} -\beta^yD_{py} )^2
         -a_1D_{+x} D_{-x}-c_1 D_{+y} D_{-y} -2b_1 D_{0x} D_{0y}
	 \bigg)u   =0  ,
      \label{eq:vpwave}
\end{equation}
in which the shift terms are treated by second order accurate
one-sided difference operators (\ref{eq:dpj}); and
(\ref{approximation2}) is related to the subluminal $W$-algorithm
(\ref{eq:subwave}) by
\begin{equation}
    V_\alpha:= W +\frac {h^2}{4} \bigg ( \alpha_1(D_{+x} D_{-x})^2 
       +\alpha_2 (D_{+y} D_{-y})^2 \bigg )u=0 ,
      \label{eq:valphawave}
\end{equation}
where Theorem 4 guarantees stability provided the inequalities
\begin{equation}\label{eq:alpha1}
\alpha_1 \ge \beta^{x^2} -a_1 =-a \, , \quad \alpha_1 \ge 0,
\end{equation}
\begin{equation}\label{eq:alpha2}
\alpha_2 \ge \beta^{y^2} -c_1 =-c \, , \quad \alpha_2 \ge 0,
\end{equation}
are satisfied.

The $V$-algorithm is related to the $W$-algorithm
by the second order accurate modification
\begin{equation}
   V = W +\frac{h^2}{4} \bigg (\beta^{x^2}(D_{+x} D_{-x})^2 
       +\beta^{y^2}(D_{+y} D_{-y})^2 \bigg )u=0 .
   \label{eq:vwwave}
\end{equation}
In the subluminal case where the $W$ and $V$ algorithms can be compared,
tests show that the $W$-algorithm has considerably better accuracy due to its
more compact stencil~\cite{babev}. Here we  carry out a set of 2D
superluminal tests to compare the performance of the superluminal algorithms
in a periodic test problem (smooth toroidal boundary conditions) where the
effect of the boundary is eliminated.  The first order accurate
$V_+$-algorithm (\ref{eq:v+wave}) is highly dissipative and much less
accurate than the second order accurate $V_p$ version (\ref{eq:vpwave}). For
these reasons, we restrict our test comparisons to the $V$, $V_p$ and
$V_\alpha$ algorithms.

The $V$-algorithm is a special case of the $V_{\alpha}$-algorithm
(\ref{eq:valphawave}) where $\alpha_1=\beta^{x^2}$ and 
$\alpha_2=\beta^{y^2}$. The accuracy of the $V_{\alpha}$-algorithm might be
expected to depend on the relative weight of the higher order terms
responsible for the stretched stencil in (\ref{eq:valphawave}). For example,
the $V_{\alpha}$-algorithm might be expected to be most accurate for the
minimum  values, $\alpha_1=(|a|-a)/2$ and $\alpha_2=(|c|-c)/2$, which are
allowed by (\ref{eq:alpha1}) and (\ref{eq:alpha2}). This is true for the case
when $a$ and $c$ are positive, for which $\alpha_1=\alpha_2=0$ and the
$V_\alpha$-algorithm reduces to the $W$-algorithm. 

When $a<0$ is negative and $c>0$, the optimal value for $\alpha_2$ remains
$0$ but the optimal value for $\alpha_1$ is not necessarily the minimum
allowed value $\alpha_1=-a$. This value would result in approximating the
$a\partial_x^2$ term in the wave operator by $aD_{0x}^2$, which decouples the
even and odd grid points. Although the optimal choice of $\alpha_1$ in this
case is not obvious, the combination $\alpha_1=\beta^{x^2}$ and $\alpha_2=0$
would give better accuracy than the $V$-algorithm. No general guidelines are
suggested by examining the truncation error in the $V_\alpha$-algorithm,
which to order $h^2$ is given by
\begin{equation}
   \tau= \frac{h^2}{12}\bigg( (a-3\alpha_1)\partial_x^4 
       + (c -3\alpha_2) \partial_y^4 
             +4b (\partial_x^3 \partial_y +\partial_x \partial_y^3 )
       +4\beta^x \partial_x^3 \partial_t +4\beta^x \partial_y^3 \partial_t
                 \bigg)u.
\label{eq:terr}
\end{equation}
Note that the values $\alpha_1=a/3$ and $\alpha_2=c/3$ correspond to the
fourth order accurate approximations to the terms $a\partial_x^2$
and $c\partial_y^2$ in the wave operator. However, these choices are not
allowed in the superluminal regime, where stability requires $\alpha_i\ge 0$.

As a test problem for comparing the accuracy of these evolution algorithms
in the superluminal regime we pick a case where both $a$ and $c$ are
negative. We consider the wave equation
\begin{equation}
  \bigg (- \partial_t^2 +4(\partial_x + \partial_y)\partial_t 
      -3\partial_x^2 - 3 \partial_y^2
       -8\partial_x \partial_y \bigg )u=0.
      \label{eq:sh2wave}
\end{equation}
which arises from a 2D version of (\ref{eq:wshift}) with shift
$\beta^x=\beta^y=2$. With this superluminal choice of shift,
there are no characteristics in the $(x>0,y>0)$ directions.
Waves propagating along the diagonal have the form 
\begin{equation}
  u = F[x+y +(4+ \sqrt{2} )t ]
       +G[x+y +(4- \sqrt{2} )t] .
\end{equation}

In our test, we simulate the solution
\begin{equation}
  u=\sin \bigg (2\pi [x+y +(4+ \sqrt{2} )t] \bigg )
   \label{eq:solPhi}
\end{equation}
in the domain $-.5\le (x,y)\le .5$, on a grid with  $N=200$ points, with
periodic boundary conditions. For this particular solution, the symmetries
$\partial_x u=\partial_y u = \partial_t u/(4+\sqrt{2})$ imply that the
truncation error (\ref{eq:terr}) has a minimum at $\alpha_1=\alpha_2
=\alpha_m$, where
\begin{equation}
       \alpha_m = \frac{13+8\sqrt{2}}{3} \approx 8.1045695.
\end{equation}

\begin{figure}[hbtp]
  \centering
  \includegraphics*[width=7cm]{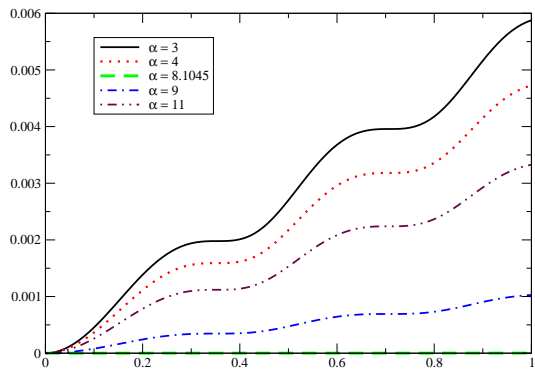}
  \caption{The $\ell_\infty$ norm of the error of the scalar field
   obtained with $V_\alpha$ algorithm on a grid of 200 points
   is plotted {\it vs} time, in the interval $0\le t\le 1$. For the value
   $\alpha_m\approx 8.1045695$, the error is barely discernible and the
   plots clearly indicate that $\alpha_m$ is the optimal value.
   The value $\alpha=4$ corresponds to the $V$-algorithm, which has 
   significantly larger error.
   }  
  \label{fig:Valp_PhiErr}
\end{figure}

Figure~\ref{fig:Valp_PhiErr} plots the $\ell_\infty$ norm of the numerical
error in the scalar field obtained in the simulation of (\ref{eq:solPhi}) by
evolving the wave equation (\ref{eq:sh2wave}) with the $V_\alpha$-algorithm,
for various values of $\alpha_1=\alpha_2=\alpha$. The error for
$\alpha=8.1045695$ is extremely small and the plots confirm that $\alpha_m$
is indeed the optimal value. The value $\alpha=4$ corresponds to the
$V$-algorithm, which gives significantly larger error. The value $\alpha=3$,
which is the smallest value allowed by stability, gives even larger error.

\begin{figure}[hbtp]
  \centering
  \includegraphics*[width=7cm]{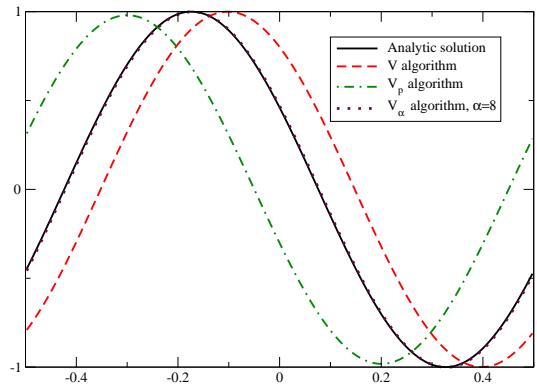}
  \caption{Snapshots of the scalar field $u(t=100,x)$ 
   obtained with the $V$, $V_p$ and $V_\alpha$
   algorithms, compared with the analytic solution. The phase error with the
   $V_p$-algorithm is larger than with the $V$-algorithm. 
   The $V_\alpha$-algorithm, for $\alpha=8$, is extremely
   accurate and barely distinguishable from the analytic solution.}  
  \label{fig:PhiSinSh2DB2}
\end{figure}

The error in Fig.~\ref{fig:Valp_PhiErr} is predominantly phase error.
Figure~\ref{fig:PhiSinSh2DB2} shows snapshots of $u(t=100,x)$ (100 crossing
times) for the simulation of (\ref{eq:solPhi}) using the $V$, $V_p$ and
$V_\alpha$ algorithms, with $\alpha=8$. The simulations are compared with the
analytical solution at $t=100$. The solution with the $V_p$- algorithm  leads
in phase while that with the $V$-algorithm lags in phase and it has slightly
better accuracy. As expected from the above error analysis, the
$V_{\alpha}$-algorithm, with $\alpha=8$, is extremely accurate.

\section{Simulation of a model 2D excision problem}
\label{sec:excision}

In this section, we simulate a simple 2D model of the excision
problem in which the inner boundary ${\cal S}$ is spacelike and
the outer boundary ${\cal T}$ is timelike, with a horizon ${\cal
H}$ in between. In the inner region between ${\cal S}$ and
${\cal H}$, since the shift is superluminal, the operator $P$ in
(\ref{shifted-system-2D-P}) is non-elliptic and both
characteristics leave the inner boundary. In the outer region
between ${\cal H}$ and ${\cal T}$, since the shift is
subluminal, the operator $P$ is elliptic and one characteristic
leaves ${\cal T}$ and the other enters ${\cal T}$. 

\begin{figure}[hbtp]
  \centering
  \includegraphics*[width=7cm]{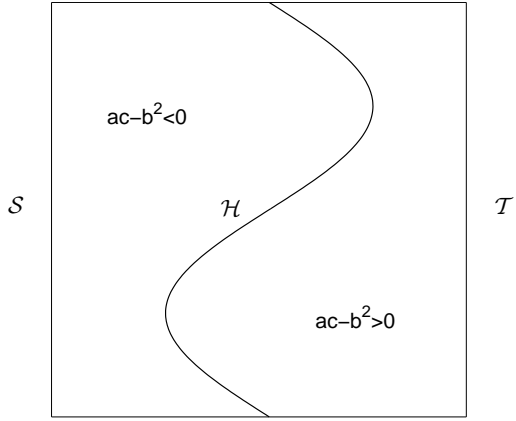}
  \caption{Computational domain with the spacelike boundary
   ${\cal S}$ on the left, the timelike
   boundary ${\cal T}$ on the right and the sinusoidal shaped
   horizon ${\cal H}$ in between. The solution is periodic
   in the vertical $y$-direction. }
  \label{fig:domain2D}
\end{figure}

To model a wave pulse propagating into a horizon, we consider
the shifted wave equation with a source term $F$,
\begin{equation}\label{numerics}
       u_{tt}=2(\beta^x u_{xt}+\beta^y u_{yt})
        +a u_{xx}+2b u_{xy}+c u_{yy}+F(x,y,t),
\end{equation}
on the spatial domain $(x,y)\in\Omega=[-2,2]\times [-2,2]$, and $t
\ge 0$. We set the coefficients $\beta^x=\beta^y=2$, $a=0.5
(x-\sin{\frac{\pi y}{2}})$, $b=0.5$ and $c=5$, for which the
problem is well-posed. The spacelike boundary ${\cal S}$ at $x=-2$,
the timelike boundary ${\cal T}$ at $x=2$ and the horizon ${\cal
H}$ are shown in Fig.~\ref{fig:domain2D}. The horizon satisfies
$ac-b^2=0$, which determines the curve
\begin{equation}
         x=0.1 + \sin\frac{\pi y}{2}.
\label{eq:horizon}
\end{equation}

For the smooth function
\begin{equation}\label{g-b}
     F(x,y,t)= {2 \over {\sigma^2}}
    \Big( -(1-2\beta^x-a)\big( \sigma-2(t+x-x_0)^2 \big)
  -4(\beta^y+b)(t+x-x_0)y+c(\sigma-2y^2) \Big)\,
    e^{-{(t+x-x_0)^2+y^2 \over {\sigma}}},
\end{equation}
the equation (\ref{numerics}) has the solution
\begin{equation}\label{solution}
    u(x,y,t)=e^{-{(t+x-x_0)^2+y^2 \over {\sigma}}},
\end{equation}
which is a left-traveling wave packet, initially centered about $x_0=0.5$
outside the horizon and propagating towards the spacelike boundary. Here we set
$\sigma = 0.05$. We uniformly discretize the spatial domain as $x_{\nu}=\nu h$
and $y_{\mu}=\mu h$ with $\nu, \mu=0,\pm 1,\dotsc,\pm N$ with the grid size $h
= {2 \over N}$.

The global simulation of the model problem in the region between ${\cal S}$ and
${\cal T}$ is carried out by combining the superluminal $V$-algorithms
established in Sec.~\ref{sec:algorithms} with the subluminal $W$-algorithm. The
spacelike boundary and the superluminal region are treated with one of the
$V$-algorithms. A region containing the timelike boundary is treated by the
$W$-algorithm.  

We consider the following three global algorithms:
\begin{itemize}

\item {\bf Algorithm 1.} The superluminal region is treated by the
$V$-algorithm (\ref{eq:vpwave}). In the subluminal region we use the
$W$-algorithm. We introduce a cut-off function $\phi$ which is 0 when $a>0$ and
$c>0$ and is 1 when $a \le 0$ or $c \le 0$. Then we use the following
approximation
$$
\phi V + (1-\phi) W = 0.
$$ 

\item {\bf Algorithm 2.} This is similar to {\bf 1}, except the superluminal
region is treated by the $V_p$-algorithm (\ref{eq:vpwave}), which is then
blended to the $W$-algorithm in the same way as in {\bf 1}.

\item {\bf Algorithm 3.} We use the $V_\alpha$-algorithm (\ref{eq:valphawave})
with $\alpha_1=(|a|-a)/2$ and $\alpha_2=(|c|-c)/2$. 

\end{itemize}

The initial data and boundary condition at $x=2$ are chosen
according to the exact solution (\ref{solution}). In the first
and third algorithms, we need two extra boundary conditions at
$\nu=-N, -N+1$. In the second algorithm, we need only one
boundary condition at $\nu=-N$. We use third order
extrapolations as the extra boundary conditions. In the
$y$-direction we use periodic boundary conditions. For the
integration in time, we use the standard 4th order Runge-Kutta
method.

\begin{figure}[hbtp]
  \centering
   \subfigure[The initial wave pulse at $t=0$.]{     
        \includegraphics[width=6cm]{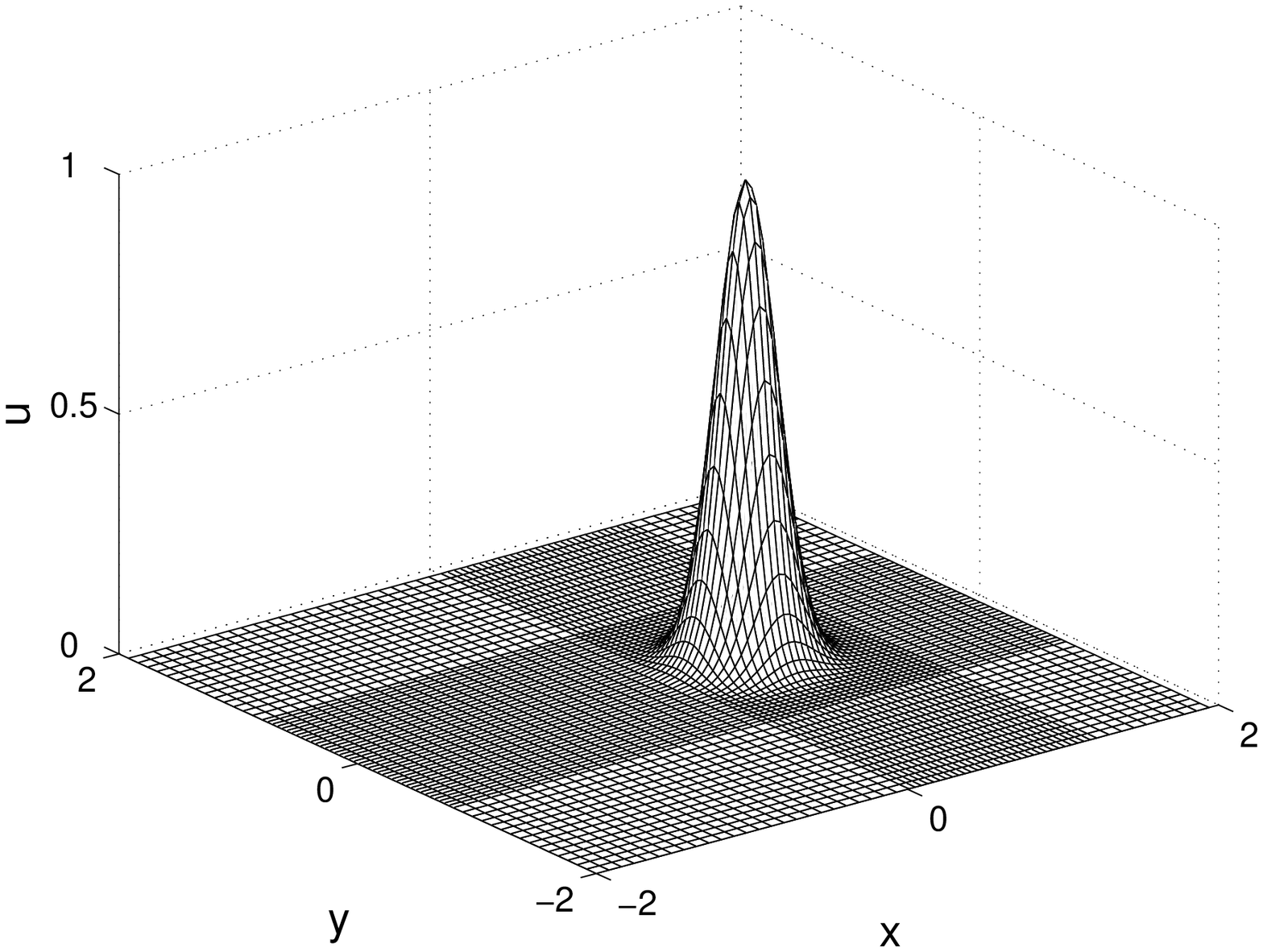}
        }
   \subfigure[The wave pulse at $t=2$.]{     
        \includegraphics[width=6cm]{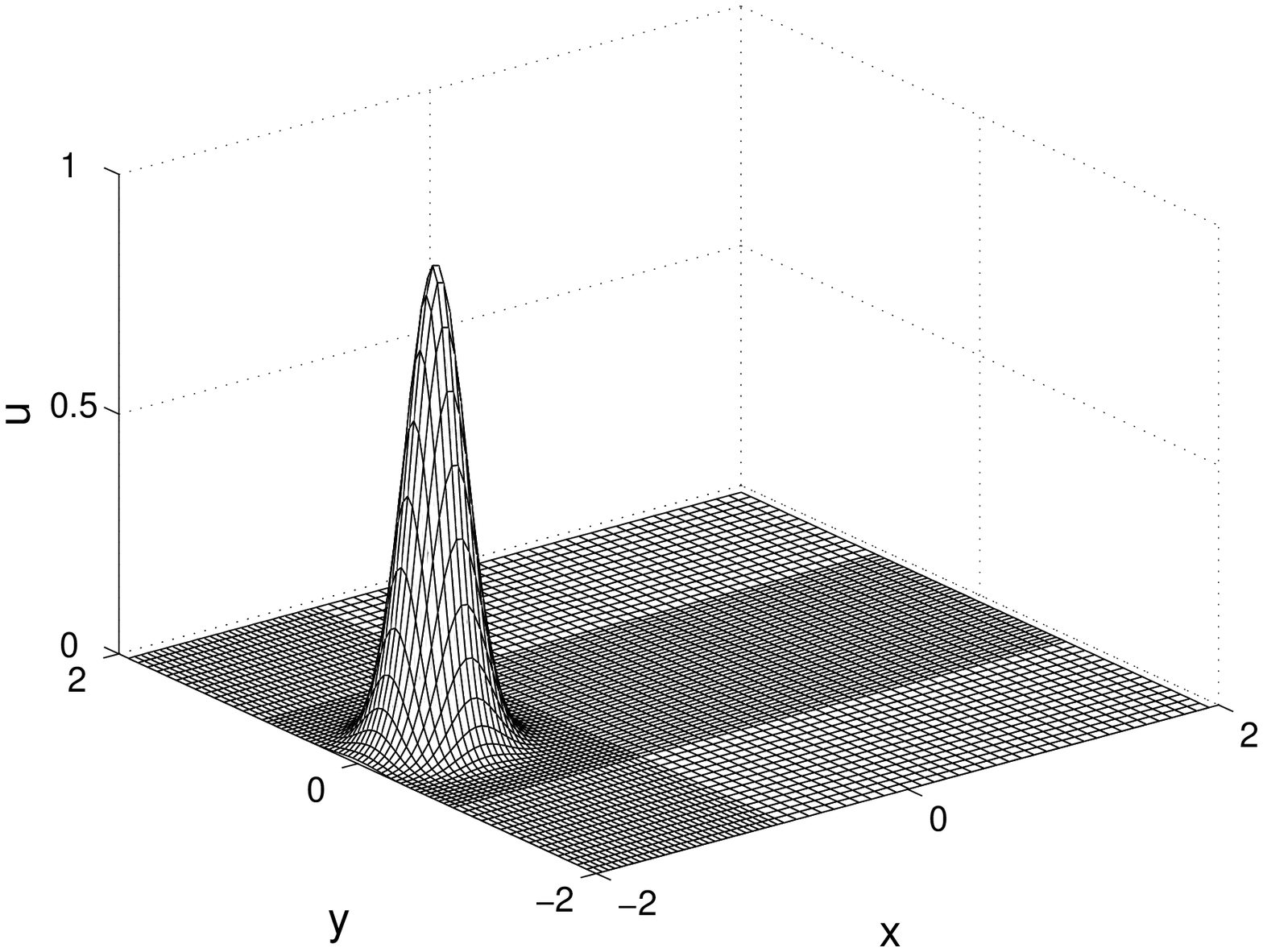}
        }
   \caption{A pulse propagating across the horizon. The left
   figure shows the initial pulse in the region outside the horizon.
   The right figure shows the pulse at a later time, after it has crossed
   the horizon and is incident on the inner spacelike boundary.}
  \label{fig:pulse}
\end{figure}

Figure~\ref{fig:pulse} shows the initial wave pulse and the pulse
at a later time $t=2$ computed by the third algorithm. 

For the gridfunction $u_{\nu \mu}(t)$ approximating $u(x_{\nu},y_{\mu},t)$,
we define the discrete norm as
\begin{equation}
   ||u_{\nu \mu}||_h^2=\sum_{\nu,\mu=-N}^{N} u_{\nu \mu}h^2,
\end{equation}
where $h=\Delta x = \Delta y$ is the gridlength. We then define
the convergence factor by
\begin{equation}\label{CONV-FACTOR}
   \mathcal{C}(t)=\log_2\left( 
      {||\mathcal{E}(t)||_h \over ||\mathcal{E}(t)||_{h/2}}
       \right),  \quad
      \mathcal{E}(t)=u(x_{\nu},y_{\mu},t)-u_{\nu \mu}(t),
\end{equation}
where $\mathcal{E}(t)$ is the error at time $t$, and
$u(x_{\nu},y_{\mu},t)$ is the exact solution computed by
(\ref{solution}).

\begin{figure}[hbtp]
\psfrag{e}{\scriptsize $||\mathcal{E}||_h$}
  \centering
  \includegraphics[width=7cm]{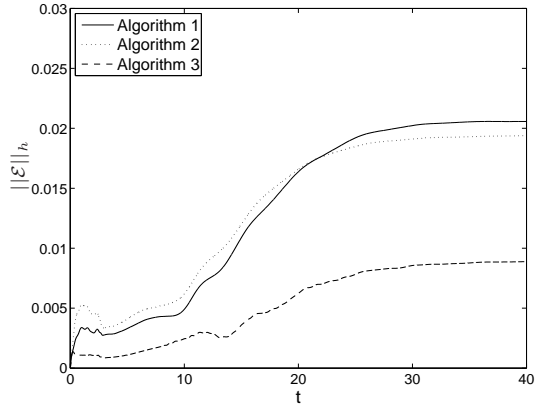}
  \caption{Norm of the error $||\mathcal{E}||_h$ versus time, with $h=0.02$.}
  \label{fig:error}
\end{figure}

Figure~\ref{fig:error} shows the norm of the error versus time for
the three algorithms with $h=0.02$ and $\Delta t=0.001$. 

\begin{figure}[hbtp]
  \centering
  \subfigure[Algorithm 1.]{
     \includegraphics[width=5cm]{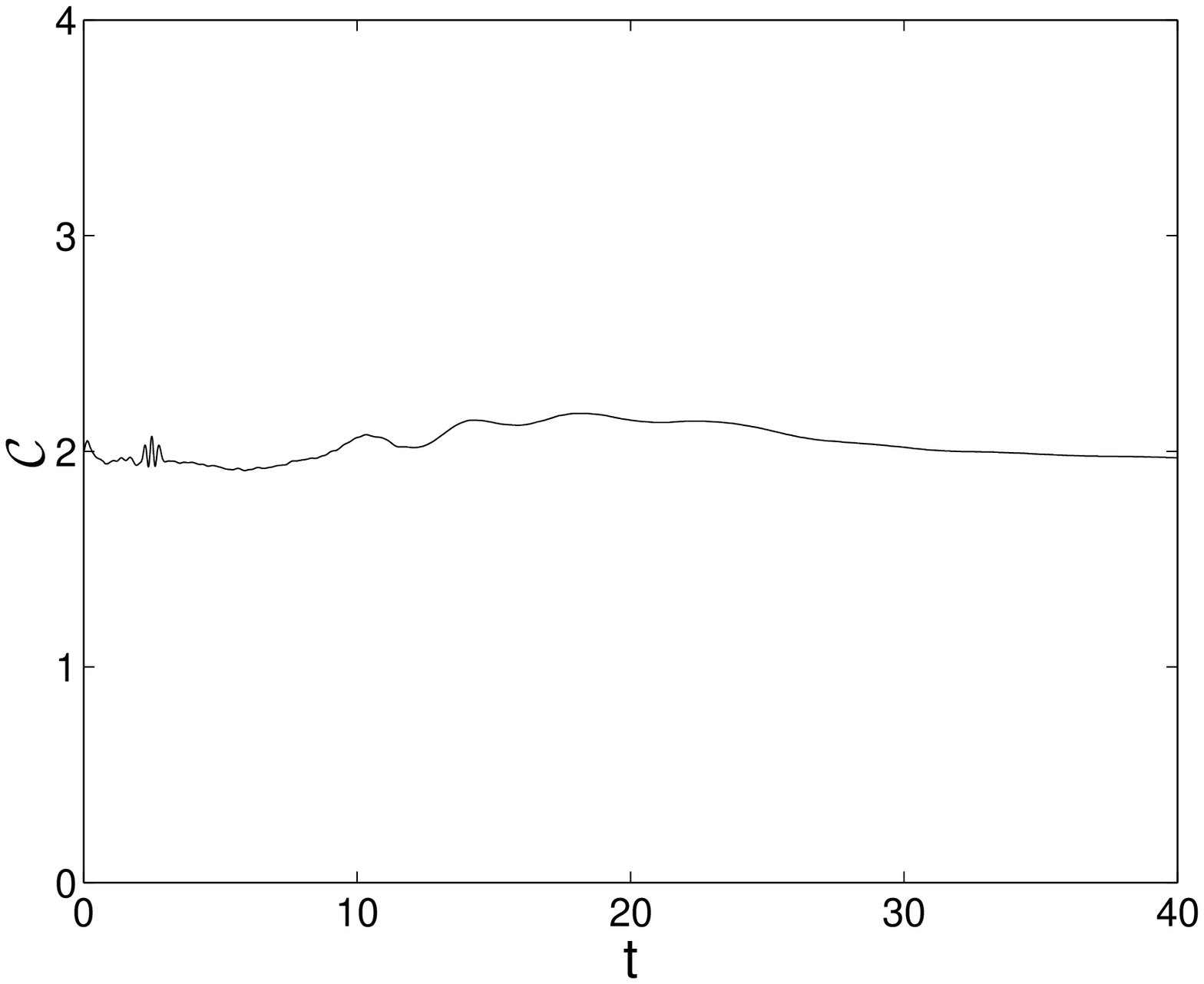}
  }
  \subfigure[Algorithm 2.]{
     \includegraphics[width=5cm]{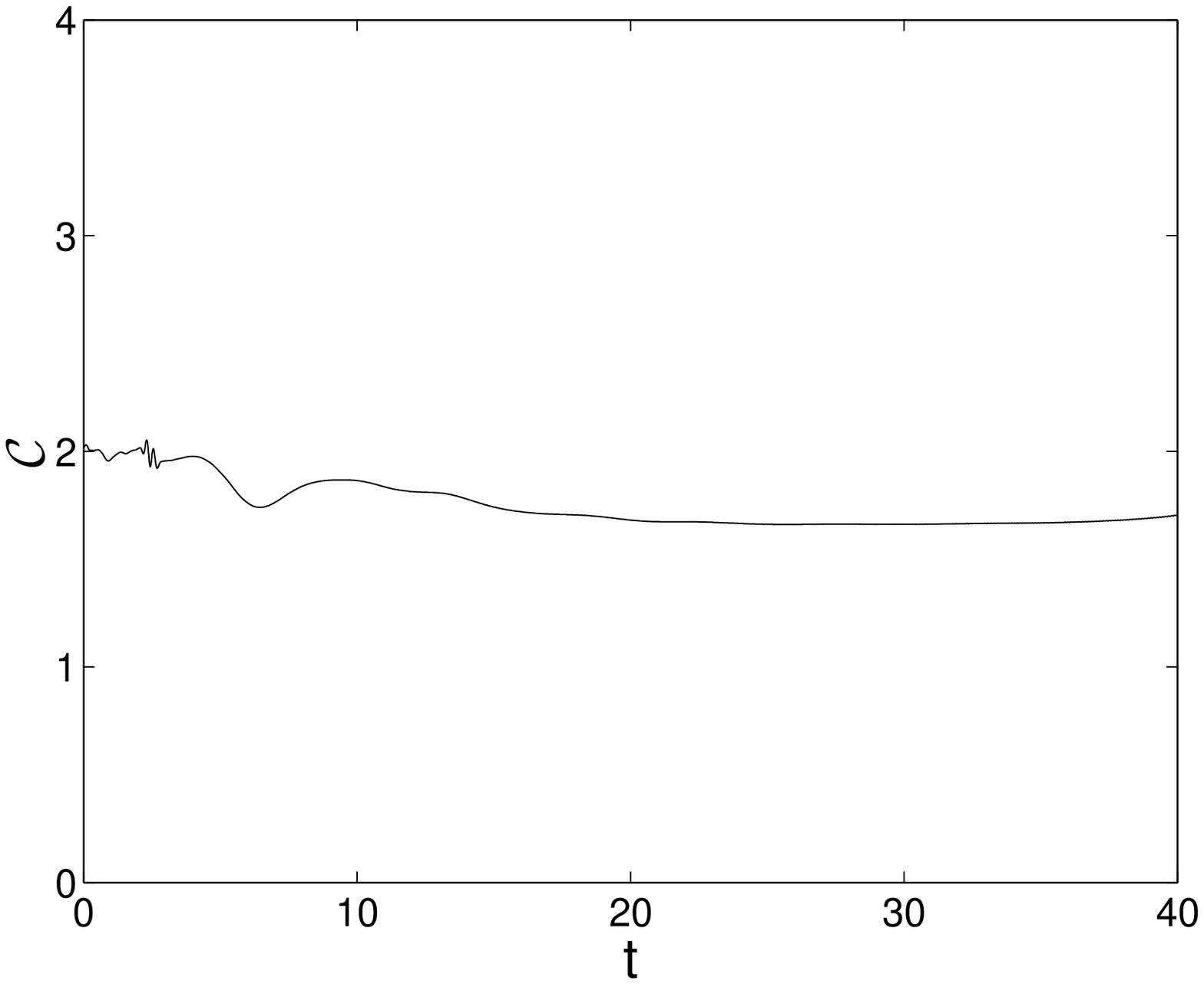}
  }
  \subfigure[Algorithm 3.]{
     \includegraphics[width=5cm]{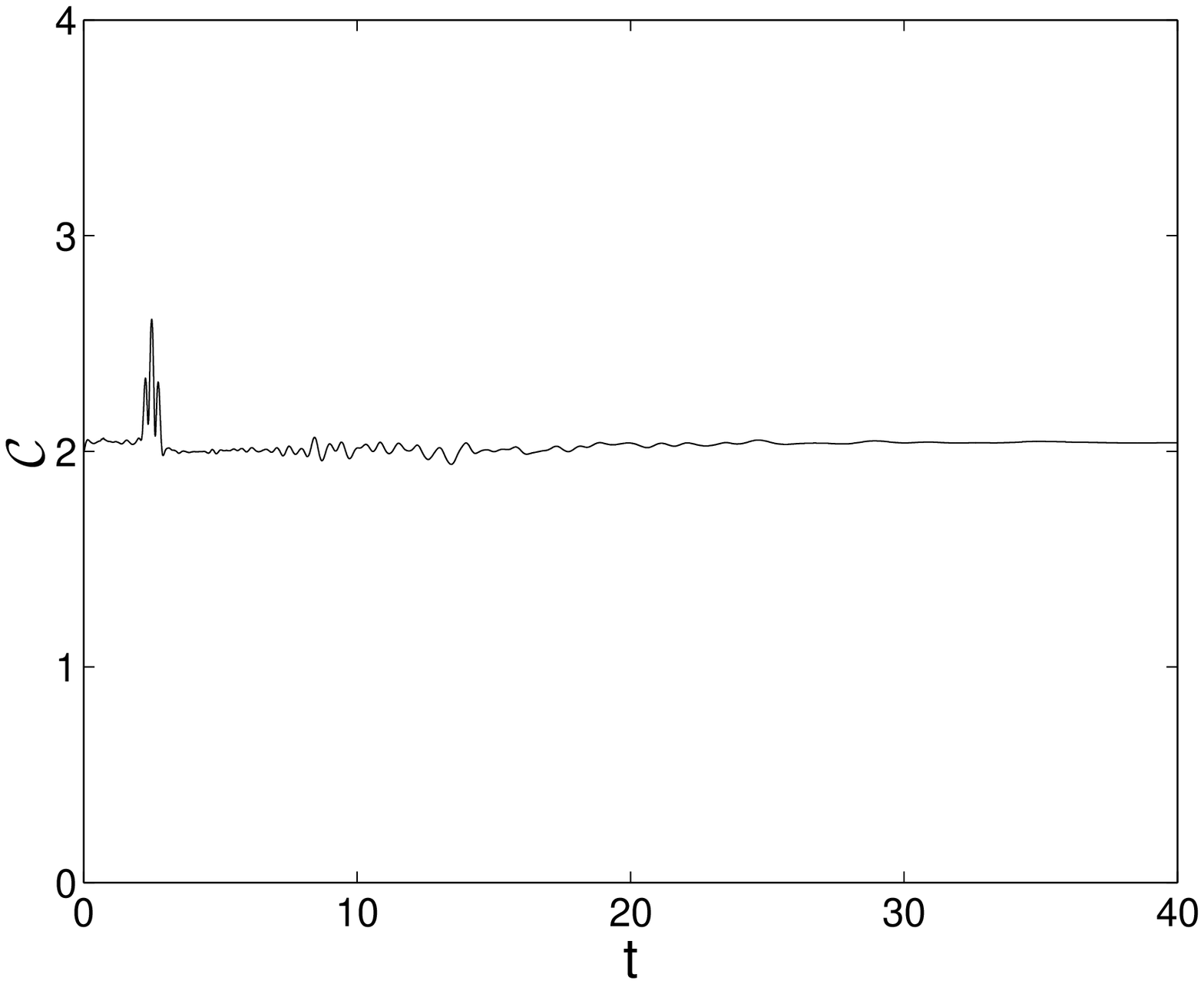}
  }
  \caption{Convergence factor $\mathcal{C}$ versus time indicates
  that the algorithms are second order accurate in space.}
  \label{fig:convergence}
\end{figure}

Figure~\ref{fig:convergence} shows the convergence factor as a function of time
for the three algorithms with $h=0.04$ and $\Delta t = 0.001$. It confirms the
second order accuracy of the algorithms in space. The jumps in the convergence
factor at about $t=2$ is a result of using third order extrapolations at the
spacelike inner boundary, while we use second order evolution algorithms. At
this time the pulse reaches the spacelike boundary and an increase in the order
of accuracy, from 2 to 3, is expected.

The third algorithm gives a better accuracy than the other two.
The second algorithm, in which we use the second order one-sided
stencil with extrapolation in one ghost point, gives a slightly smaller error than the first algorithm, in which the second order centered stencil with extrapolation in two ghost points is used.  

\begin{acknowledgments}

This work was supported by the National Science Foundation under
grant PH-0244673 to the University of Pittsburgh. We have used
computing resources of the Pittsburgh Supercomputing Center and
of the National Energy Research Scientific Computing Center; and
we have benefited from the use of the Cactus Computational
Toolkit (http://www.cactuscode.org).

\end{acknowledgments}

\end{document}